\documentclass[article]{aastex631}

\usepackage{graphicx}

\begin{document}

\title{Titan's Fluvial \& Lacustrine Landscapes}

\author[0000-0002-4578-1694]{Samuel P.D. Birch}
\affiliation{Department of Earth, Atmospheric and Planetary Science, Massachusetts Institute of Technology, Cambridge, MA 02139, USA}

\author[0000-0001-6397-2630]{Alexander G. Hayes}
\affiliation{Cornell Center for Astrophysics and Planetary Sciences, Cornell University, Ithaca, NY 14853, USA}

\author[0000-0002-0404-8701]{J. Taylor Perron}
\affiliation{Department of Earth, Atmospheric and Planetary Science, Massachusetts Institute of Technology, Cambridge, MA 02139, USA}

\section{Introduction}
\begin{figure}[h]
\centering
\includegraphics[width=\textwidth]{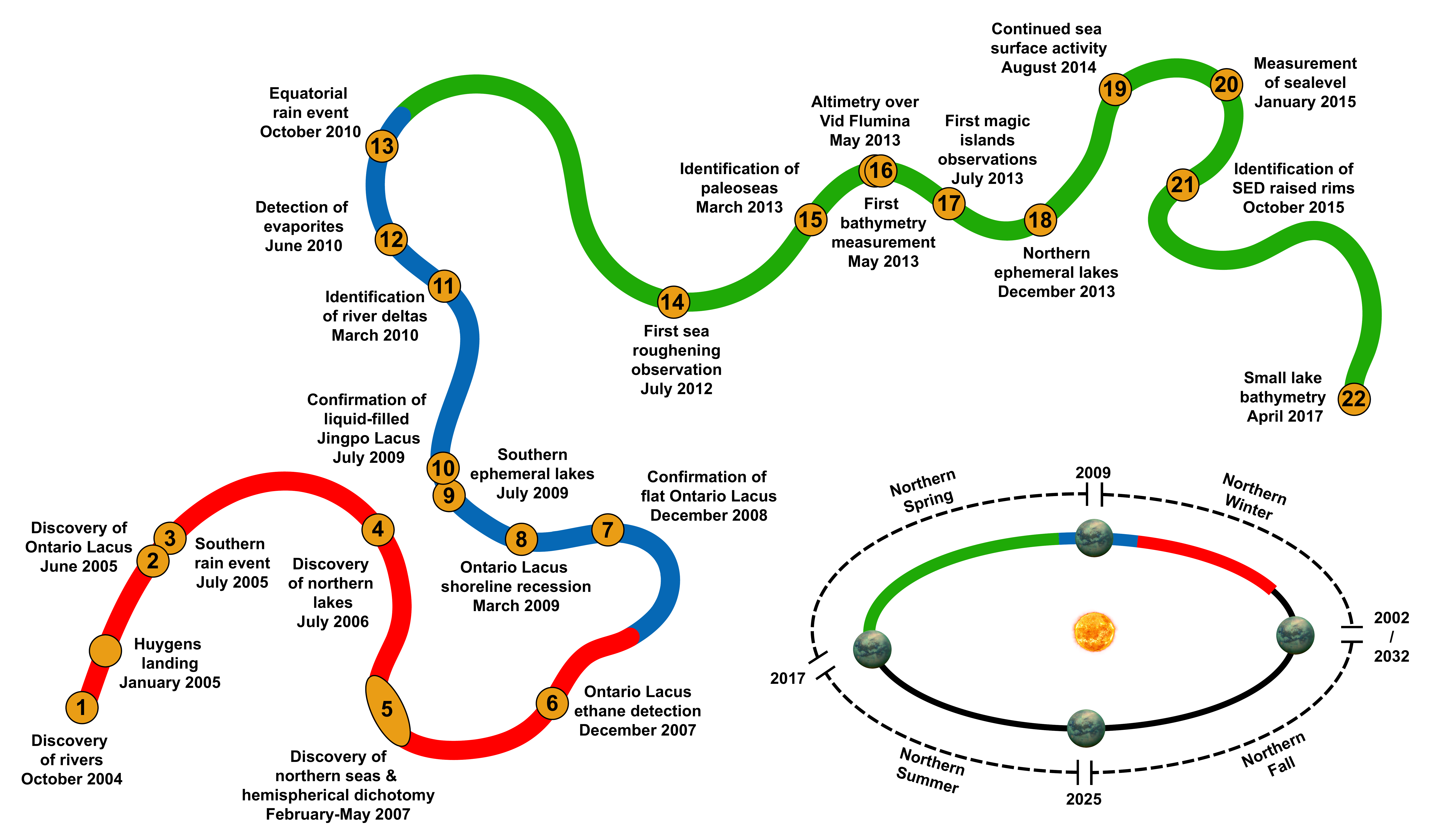}
\caption{\textbf{Timeline of \textit{Cassini}'s and \textit{Huygens}' fluvial \& lacustrine discoveries.}  The red, blue, and green colors denote \textit{Cassini}'s primary, equinox and solstice missions, which took place over less than one half of a Titan year (graphic inspired by Ralph Lorenz). The events listed are described in: (1) \citet{lorenz_fluvial_2008}. (2/3) \citet{turtle_cassini_2009}. (4) \citet{stofan_lakes_2007}. (5) \citet{aharonson_asymmetric_2009}. (6) \citet{brown_identification_2008}. (7) \citet{wye_smoothness_2009} and \citet{barnes_shoreline_2009}. (8) \citet{turtle_shoreline_2011}. (9) \citet{hayes_transient_2011}. (10) \citet{stephan_specular_2010}. (11) \citet{wall_active_2010}. (12) \citet{barnes_organic_2011}. (13) \citet{turtle_rapid_2011}. (14) \citet{barnes_cassinivims_2014}. (15) \citet{wood_paleosea_2013} and \citet{birch_morphological_2018}. (16) \citet{poggiali_liquid-filled_2016} and \citet{mastrogiuseppe_bathymetry_2014}. (17) \citet{hofgartner_transient_2014}. (18) \citet{mackenzie_case_2019}. (19) \citet{hofgartner_titans_2016} among others. (20) \citet{hayes_topographic_2017}. (21) \citet{birch_raised_2019}. (22) \citet{mastrogiuseppe_deep_2019}.}
\label{Discovery_timeline}
\end{figure}

The idea of rivers, lakes and seas on Titan, like on Mars \citep{crossley_canali_2000}, was first popularized within the realm of Science Fiction. In `The Sirens of Titan' (1959), Kurt Vonnegut describes Titan as having three seas of similar size to Lake Michigan, connected via valley networks to a series of smaller lakes and ponds. As it turns out, Vonnegut’s description was strikingly similar to reality. \textit{Cassini} revealed that Titan, like  Earth, has vast bodies of standing liquid on its surface today, with rivers actively modifying its landscapes and interacting directly with the atmosphere (Figure \ref{Discovery_timeline}). This makes Titan and the Earth unique in the solar system – the only two such worlds where active rivers, lakes, and seas are found.

While \textit{Cassini} has revealed Titan to be a hydrologically active world with landscapes shaped by processes strikingly similar to their terrestrial counterparts \citep{lunine_methane_2008,hayes_lakes_2016,mitchell_climate_2016,mackenzie_titan_2021}, we have barely scratched the surface of understanding the complex coupling between Titan's landscapes and climate. Understanding Titan's landscapes and climate also goes far beyond just developing a better understanding of this one icy moon of Saturn. On Earth, just like Titan, the origins of our many landscapes remain stubbornly enigmatic (e.g., What sets a river channel's width? $-$ \citet{dunne_what_2020}), a problem exacerbated by the simple fact that the landscapes we observe today result from a convolution of multiple processes, all of which act over timescales that can differ by orders of magnitude \citep{perron_climate_2017}. Despite these challenges, significant progress has been made by seeking out natural experiments, wherein one variable can be controlled and the response of the landscape can be observed and quantitatively modeled (e.g., \citet{perron_formation_2009,perron_root_2012,ferrier_covariation_2013,perron_climate_2017,huppert_influence_2020}). Titan is an active natural experiment on a global scale! Detailed study of its surface can provide the strictest tests of the universality of physical processes we are still seeking explanation for here on Earth, tests not possible to undertake anywhere else. 
 
For example, on Earth geologic evidence suggests that there has been liquid water for most of the last few billion years \citep{mojzsis_oxygen-isotope_2001,wilde_evidence_2001}. How does a planet maintain a climate that remains so close to its primary volatile's phase transition for so long, and what regulatory processes and feedbacks maintain it? On Earth, and perhaps on some rocky exoplanets, active tectonism continually creates new relief that exhumes new rocks that can chemically weather, a process that may buffer the planetary climate for millennia \citep{berner_carbonate-silicate_1983,kasting_habitable_1993}. Mars may have never had such a planetary thermostat, and it is unclear how general such a process is within the solar system and beyond. On Titan we do not yet know how long its methane-based hydrologic cycle has been active and what weathering processes may be at work. But it must have been long enough to modify its surface to the degree observed, and it is unlikely that we happen to be observing Titan at a special time. By looking at Titan's hydrologic landforms, we can therefore gain an understanding of the vigor and longevity of its activity, knowledge that should help reveal new information about how planetary climates can maintain a volatile cycle, how planetary topography is maintained through time, and the nature of Titan's alien, astrobiologically intriguing, surface materials.

In this chapter we begin with a review of Titan's fluvial and lacustrine landscapes as observed with \textit{Cassini} remote sensing data, and what the many discoveries have revealed about Titan's surface materials and climate. Yet \textit{Cassini} remote sensing data are coarse, topographic data are largely lacking \citep{corlies_titans_2017}, and the absence of in situ field measurements means we have little understanding of what the surface is composed of. At present, our knowledge of Titan's hydrology is comparable to that of Mars in the 1970's during the Viking era (Figure \ref{Huygens-Cassini}). Fortunately, the coming decades promise many new and exciting discoveries that can be achieved through Earth-based experiments, numerical modeling, and a continued commitment to the exploration of Titan by future missions, including both \textit{Dragonfly} \citep{barnes_science_2021} and orbiting assets \citep{mackenzie_titan_2021}. We therefore close the chapter with a discussion about what can be done with the current \textit{Cassini} data and how new data, from both \textit{Dragonfly} and a potential future orbiter, would allow us to leverage Titan to help solve some of the largest problems both here on Earth and on hydrologic planets and exoplanets more generally.
\newline
 
\section{Titan's Fluvial Landscapes}
\begin{figure} [h]
\centering
\includegraphics[width=\textwidth]{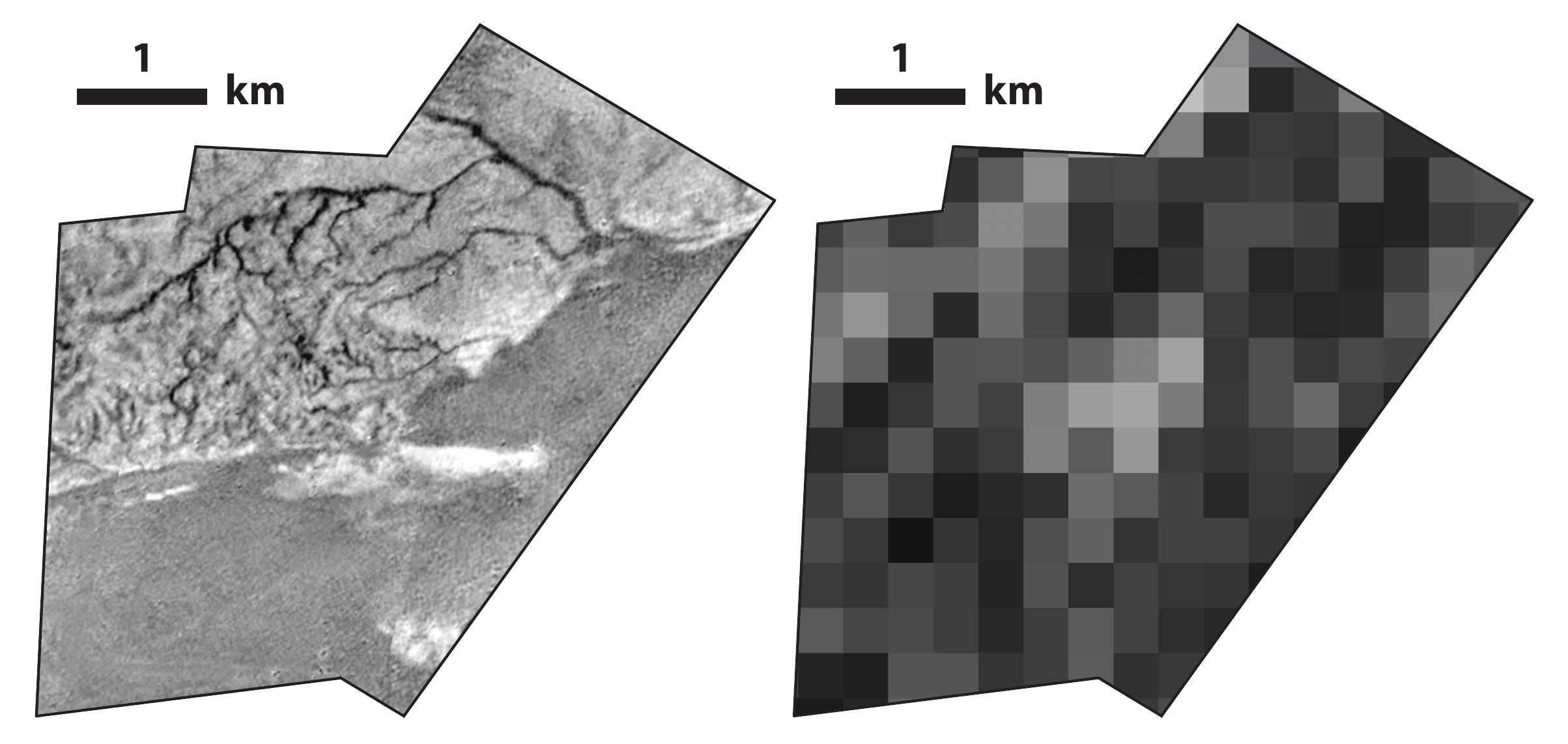}
\caption{\textbf{Comparison between \textit{Cassini} and \textit{Huygens} imaging.} Left: \textit{Huygens} DISR mosaic of a dendritic fluvial network near its landing site. Right: How \textit{Cassini}'s SAR instrument saw the exact same surface, highlighting just how little we have observed of Titan in detail.}
\label{Huygens-Cassini}
\end{figure}

\textit{Cassini} Synthetic Aperture Radar (SAR), Visual and Infrared Mapping Spectrometer (VIMS) and Imaging Science Subsystem (ISS) image and topographic data have been used by multiple authors to study and map the distribution of fluvial valley networks across the surface of Titan \citep{collins_relative_2005,perron_valley_2006,barnes_near-infrared_2007,lorenz_fluvial_2008,jaumann_fluvial_2008,cartwright_channel_2011,langhans_titans_2012,burr_fluvial_2013,burr_morphology_2013,miller_fluvial_2021,birch_geometry_2022}. The majority of Titan’s observable valley networks are found in the polar regions around the largest seas, and within the mountainous Xanadu region \citep{radebaugh_regional_2011} near Titan’s equator (Figure \ref{Distribution}; \citet{miller_fluvial_2021}). Yet due to the limited resolution and coverage of \textit{Cassini} SAR images ($\sim$20$\%$ of the surface at $<$350 meters and $\sim$40$\%$ at $<$1 kilometers \citep{hayes_post-cassini_2018}), the mapping is likely to be incomplete for valley networks $<$700 meters in width \citep{miller_fluvial_2021}. That the surface is highly dissected at the single location observed at finer resolution by \textit{Huygens} (Figure \ref{Huygens-Cassini}) suggests that there may be abundant rivers below \textit{Cassini}’s resolution across the whole of Titan, a hypothesis testable with \textit{Dragonfly}’s mid-2030’s arrival at Selk crater. 
\newline

\subsection{Liquid-Filled Polar Valley Networks}
\begin{figure}[h]
\centering
\includegraphics[width=\textwidth]{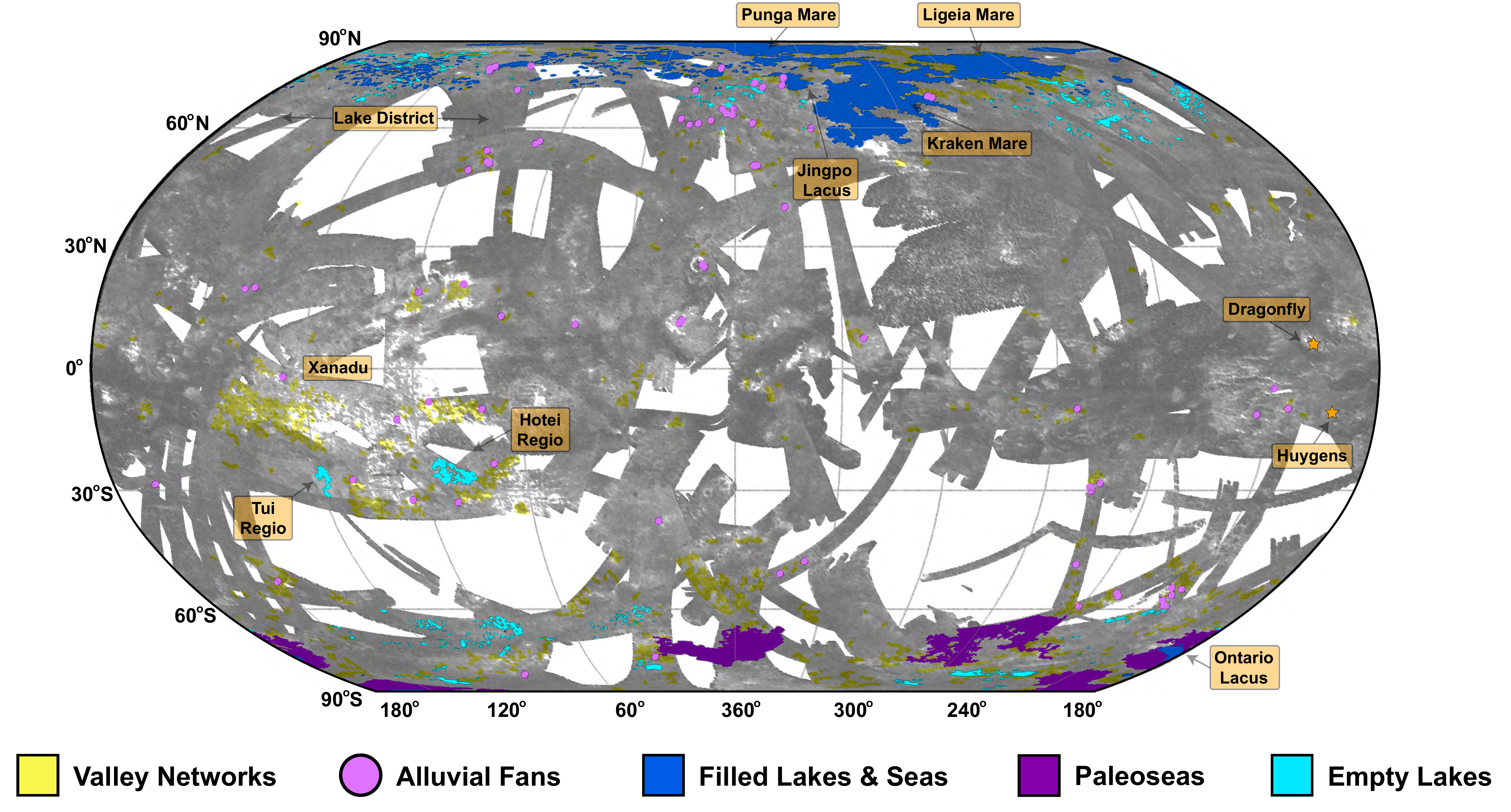}
\caption{\textbf{Distribution of fluvial and lacustrine features on Titan.} Most of Titan's valley networks, lakes, and seas are concentrated at polar latitudes, though valley networks are comparably more globally distributed \citep{miller_fluvial_2021}. Alluvial fans are primarily found at Titan's mid-latitudes \citep{birch_alluvial_2016}. At both poles, Titan's largest lakes, seas, and paleoseas concentrate on the eastern hemisphere, while most of the smaller lakes are on the opposite hemisphere \citep{birch_geomorphologic_2017}. The largest cluster of small lakes are in Titan's `Lake District', a region elevated above the seas by 100's of meters. Major regions and features are labeled, along with approximated locations of the \textit{Huygens} lander and the \textit{Dragonfly} landing site at Selk Crater.}
\label{Distribution}
\end{figure}

Titan’s largest valley networks, both in terms of apparent width and network length, are located within its polar regions (Figure \ref{Distribution}). This distribution is slightly biased due to the known presence of fluid within the polar networks \citep{miller_fluvial_2021}, likely methane-nitrogen mixtures reflective of Titan’s rain \citep{graves_rain_2008}, providing more contrast with the brighter surroundings and thereby making them more easily observed \citep{miller_fluvial_2021} than the dry, or mostly dry, equatorial networks \citep{mitchell_climate_2016}. Within the polar regions, valley networks cluster on the eastern hemispheres of both poles (Figure \ref{Distribution}), draining into the large filled and empty seas \citep{birch_geomorphologic_2017,miller_fluvial_2021}. Few fluvial features are observed in the vicinity of the smaller lakes \citep{hayes_lakes_2016,birch_geomorphologic_2017,hayes_post-cassini_2018}, suggesting that the conditions in such regions are not conducive to the formation of valley networks. 

The highest density of valley networks observed anywhere on Titan are around the liquid-filled northern seas. Valleys appear to intersect the shorelines of the seas at many points along their perimeters when sufficient resolution is available. The morphology of these networks are closely analogous to flooded river valleys on Earth, with very wide channel mouths, and liquids extending far up individual valleys into the surrounding headlands (Figure \ref{Titan_rivers}b). \textit{Cassini} altimeter observations of the liquid surface elevations of one such network, Vid Flumina, show the liquid elevations to be consistent with sea level to within the centimeter-precision of the altimeter over liquids \citep{poggiali_liquid-filled_2016}. These observations are consistent with the morphological observations and with the early predictions of liquid cycling between the poles as a consequence of orbital forcing  \citep{aharonson_asymmetric_2009,lora_simulations_2014}. However, whether fluids cycle between Titan's poles in meaningful quantities remains up for debate, as more recent climate modeling that includes surface topography struggles to generate multiple large variations in the liquid inventories between the poles (\citet{lora_topographic_2022}, see Chapter 8). Nevertheless, it is difficult to reconcile morphological observations without the influence of rising northern sea levels, at least in the recent past.  

The polar valley networks also display great morphologic diversity. Some networks, such as Vid Flumina near Ligeia Mare (Figure \ref{Titan_rivers}a), appear dendritic \citep{burr_fluvial_2013,poggiali_liquid-filled_2016,miller_fluvial_2021}, with speculative evidence of drainage capture near its margins (Figure \ref{Titan_rivers}a). Vid Flumina is also incised with steep-sided ($>$30$^{\circ}$; \citet{poggiali_liquid-filled_2016}) walls, hinting that topographic and/or base level variations may be recent/ongoing. Valley networks near Punga Mare are distinct in that they resemble sapping channels (Figure \ref{Titan_rivers}i), perhaps evidencing uniform/dissolution erosion (\citet{palermo_coasts_2022}, in prep). That such features occur only at a single point along a sea margin suggests that surface compositional variations, not detected by VIMS or ISS, may be present and make the formation of such networks favorable. Many valley networks around Kraken Mare and Ligeia Mare are also short in their spatial length, suggesting more limited flooding and/or steeper, smaller catchments. Despite the general lack of topographic data \citep{corlies_titans_2017}, morphological indicators such as drainage basin shapes \citep{miller_fluvial_2021}, network morphology \citep{black_estimating_2012}, and shoreline sinuosity \citep{tewelde_estimates_2013} all suggest that these polar valley networks have experienced some limited erosional exhumation of the surrounding landscape \citep{black_estimating_2012}. Collectively, Titan's fluvial landscapes indicate that its surface is still responding to ongoing climate variations, perhaps exacerbated by a global decline in methane inventory (\citet{hayes_lakes_2016}, see Chapter 8). 

Numerous liquid-filled valleys, such as Xanthus Flumen along the northern shoreline of Ligeia Mare (Figure \ref{Titan_rivers}g), extend tens of kilometers off shore, along the seafloor. Others like Kokytos Flumina flow parallel to the shoreline for $>$50 kilometers before ultimately entering the sea (Figure \ref{Titan_rivers}b), perhaps reflecting recent base level rise. Similar features drain into Ligeia Mare at two other points along its southern coast, and into both Kraken Mare and Jingpo Lacus along their northern coastlines, suggesting a consistent pattern (e.g, Figure \ref{Titan_rivers}b/f). Jingpo Lacus is particularly distinct, as the river valley appears to enter, exit and then re-enter the lake along its traverse (Figure \ref{Titan_rivers}f). That all such valleys remain significantly darker than their surroundings when traversing the seafloor for such distances implies either finer grained material coats only the submerged channel beds, that the valleys are significantly incised into their seafloor, or their composition is different due to different fluids and/or dissolved/suspended solids. The former is unlikely, given that deposition of fine-grained material under Titan’s current climate at these positions offshore should be uniform across both the surrounding plain and channel bed. Sea level rise may be able to explain these observations, though the valleys must be highly incised (100's of meters, \citet{mastrogiuseppe_deep_2019}) and starved of infilling deposition, both of which together are not likely. Instead, these features may point to some degree of continued incision/excavation, perhaps by hyperpycnal flows (e.g., turbidity currents) along the seafloor that are capable of outpacing any infilling under Titan’s current climate. Such density currents could arise due to the theorized large temperature-dependent density and viscosity variations of Titan’s ternary fluids \citep{steckloff_stratification_2020}, with a methane-nitrogen fluid being negatively buoyant compared to a slightly more ethane-rich sea \citep{steckloff_stratification_2020}. These flows could also be more absorptive to \textit{Cassini}'s RADAR, especially if they carry higher sediment loads, explaining their darker appearance. Though the exact dynamics and implications require detailed numerical and laboratory investigations to test the feasibility of such a process, the ubiquity of these features suggests that Titan’s fluids may frequently produce phenomena not as regularly encountered on Earth, where our single fluid, water, only exhibits relatively minor changes in its density: $\sim$1\% maximum, compared to up to $\sim$20\% on Titan \citep{steckloff_stratification_2020}.

At Titan’s south, a similar number of valley networks are observed compared to the north \citep{miller_fluvial_2021}, and are clustered around the large, though empty \citep{miller_fluvial_2021} basins (Figure \ref{Distribution}). This distribution suggests that the overall conditions necessary for channel formation are comparable between the poles \citep{birch_geomorphologic_2017}. The morphology of networks at the south are also similarly diverse to the northern networks. For example, Saraswati Flumen appears to sinuously flow across a low sloping plain (Figure \ref{Titan_rivers}e), thought to be a former seafloor composed of unconsolidated material \citep{birch_morphological_2018}. This would make Saraswati Flumen a potentially alluvial channel (\citet{birch_geometry_2022}, see below). Near Titan’s South Pole, there are also numerous straight, parallel networks that appear incised into a block tilted toward the interior of the Romo Basin (Figure \ref{Titan_rivers}d). These were used to argue for base level lowering and incision in the geologically recent past \citep{hayes_lakes_2016,birch_morphological_2018}. Finally, draining into Rossak Planitia is Celadon Flumina (Figure \ref{Titan_rivers}c), a potentially meandering valley \citep{malaska_high-volume_2011}. Though only a small fraction of the valley was ever imaged by \textit{Cassini}, the portions that were imaged are remarkably similar to meandering channels on Earth. If we are observing the channel form, and not the broader valley simply working its way around undulating topography, the presence of such features would suggest that either cohesive materials may be present on Titan’s surface and acting to stabilize the channel banks, or that yet other mechanisms may be capable of producing a meandering channel \citep{howard_how_2009}.
\newline

\subsection{Equatorial \& Mid-Latitude Valley Networks}
\begin{figure}[h]
\centering
\includegraphics[width=\textwidth]{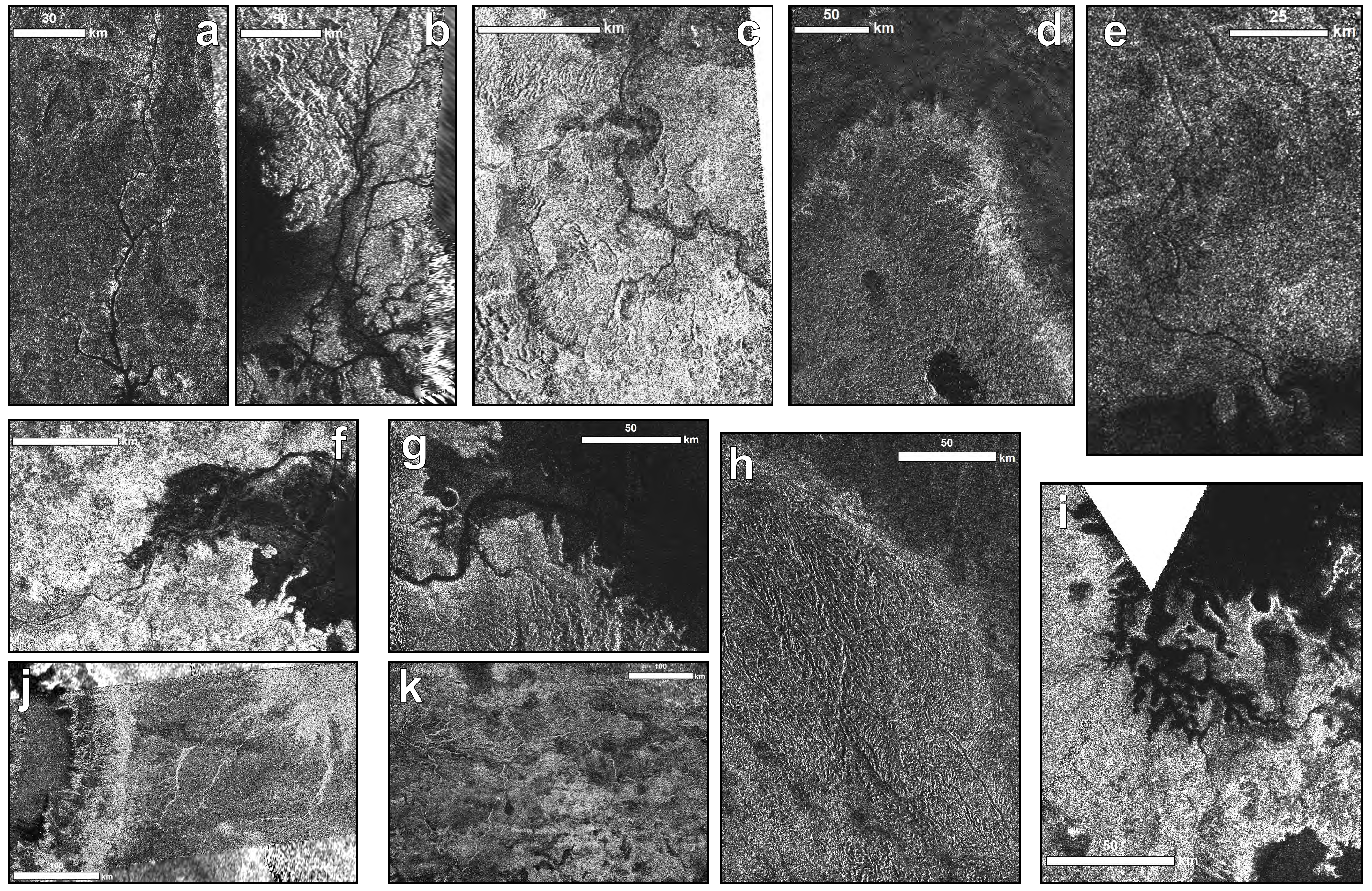}
\caption{\textbf{Diversity of fluvially-sculpted landscapes on Titan.} (a) The dendritic and incised Vid Flumina draining into Ligeia Mare. (b) Kokytos Flumina, twin networks flowing parallel to the shoreline of Ligeia Mare. (c) The sinuous Celadon Flumina at Titan's south. (d) An unnamed valley network eroding along the Romo basin perimeter. (e) Saraswati Flumen, terminating at Titan's only two delta-like features along Ontario Lacus' shoreline. (f) An unnamed valley network that appears to drain into, out of, and back into Jingpo Lacus. (g) Xanthus Flumina draining into Ligeia Mare and extending 10's of kilometers offshore along the seafloor. (h) Disconnected networks within the Lankiveil labyrinth terrains, highlighting their distinct complexity. (i) An unnamed network draining into Punga Mare that hints at evidence of sapping and/or uniform erosion. (j) Elivagar Flumina, a braided network draining off the eroded rim of Menrva crater into a broad alluvial fan complex. (k) Vast rectangular networks crossing the Xanadu region on Titan.}
\label{Titan_rivers}
\end{figure}

Across the lower latitudes of Titan, valley networks and deposits are observed in abundance, though less so than at the poles (Figure \ref{Distribution}). In particular, the Xanadu region near Titan’s equator is highly mountainous \citep{radebaugh_regional_2011} and contains the vast majority of Titan’s low-latitude valley networks (Figure \ref{Distribution}; \citet{miller_fluvial_2021}). That few networks are seen outside of Xanadu at \textit{Cassini}'s kilometer-scale resolution, suggests that sufficient slopes are unavailable outside this mountainous region, that many channels in low-lying areas have been buried as a result of aeolian sediment deposition \citep{lopes_nature_2016,malaska_geomorphological_2016,brossier_geological_2018,lopes_global_2020}, or that such channels are small and difficult to differentiate from their surroundings in \textit{Cassini} data. Fluvial valley networks are also observed in Titan’s labyrinth terrains (Figure \ref{Titan_rivers}h), with their organization and spatial extents used to classify these terrains \citep{malaska_labyrinth_2020}. The overall lack of integrated networks traversing these terrains, despite having some of the highest relief on Titan, was used to argue for karst-like erosion \citep{malaska_labyrinth_2020}. It is not possible, however, to discount the interpretation that the disconnected nature and presence of apparently closed valleys are related to \textit{Cassini}’s coarse resolution and lack of fine scale topographic data. Alternative non-karst origins for these features should therefore be investigated until new remote sensing data become available. 

The majority of the equatorial valley networks are bright in SAR images compared to their surroundings (Figure \ref{Titan_rivers}j/k) \citep{burr_fluvial_2013,burr_morphology_2013,miller_fluvial_2021}, suggesting either centimeter-sized gravel comprises their now dry, or mostly dry, channel beds \citep{le_gall_radar-bright_2010}, or that the channels reside within broader valleys that have roughness elements along their walls, exposed due to the relative lack of fluid fill. Many of the bright valleys also appear rectangular in their planform morphology (Figure \ref{Titan_rivers}k), suggesting a degree of tectonic control \citep{burr_morphology_2013}. The change in brightness of a few of these valley networks has been used as evidence of downstream fining due to size-selective transport \citep{maue_rapid_2022}, though the abrasion rates of such materials have only begun to be studied \citep{maue_rapid_2022}, and whether the channel brightness is separable from the valley brightness remains unknown. 
 
Images from the \textit{Huygens} lander during its descent also show a fluvially incised \citep{tomasko_rain_2005,daudon_new_2020} surface at scales far below \textit{Cassini}’s resolution (Figure \ref{Huygens-Cassini}). Within a small hummock in the broader Shangri-La dune field near Titan's equator (Figure \ref{Distribution}) are both a dendritic network and a more rectangular network \citep{tomasko_rain_2005}. The high drainage density of the dendritic network, with valleys extending seemingly up to the divide, was used to infer formation via precipitative-driven runoff, with moderate rates of rainfall required \citep{perron_valley_2006}. Meanwhile, the more rectangular networks were argued to be evidence of sapping erosion \citep{soderblom_topography_2007}. However, both due to coarse imaging resolution over this network compared to the dendritic channels \citep{tomasko_rain_2005,soderblom_topography_2007}, and other processes being capable of producing similar sapping-like forms \citep{lamb_can_2006,lamb_formation_2007,lamb_formation_2008},  interpretations of the rectangular network remain tenuous without images of higher-order tributaries \citep{malin_groundwater_1999}. 
\newline

\subsection{Fluvial Deposits}
Channels on Titan are likely transporting sediment, either currently or in the recent past. Gravel-sized sediment was observed at the \textit{Huygens} landing site \citep{tomasko_rain_2005}, and significant volumes of sediment are expected to deposit from the atmosphere as well (e.g., \citet{krasnopolsky_photochemical_2010}). Deposits at the end of Titan's valley networks should therefore be expected. 

Indeed, alluvial fans (Figure \ref{Titan_rivers}j) are found across Titan (Figure \ref{Distribution}; \citet{birch_alluvial_2016,radebaugh_alluvial_2016}, with a strong clustering near its mid-latitudes \citep{birch_alluvial_2016}. These fans also originate from drainage basins with greater slopes \citep{radebaugh_alluvial_2016}, suggesting similar dynamics to large megafans on Earth. VIMS observations show that the fans often appear `blue' \citep{brossier_geological_2018}, interpreted to indicate a relative enrichment in  water ice \citep{barnes_near-infrared_2007}. The global distribution of these features are consistent with global climate models, which predict that the largest discharges from rainfall occur at these same latitudes \citep{faulk_regional_2017}. These correlations suggest that these deposits are forming actively in Titan’s current climate. Their relatively small size compared to fans on Mars and Earth, however, were used to infer that Titan’s mid-latitude valley networks may be armored by coarse gravel \citep{birch_alluvial_2016}, slowing the overall lowering of Titan’s relief \citep{howard_formation_2016}.

Surprisingly, most of the larger valley networks on Titan leave no observable deposits near their termini. This is the case both for the polar seas, where deltas are seemingly rare \citep{birch_detection_2022} outside of Ontario Lacus \citep{wall_active_2010}, and for the large multi-kilometer wide bright valley networks at the equator (Figure \ref{Titan_rivers}k). Whether Titan’s rivers are simply inefficient at forming deposits, or whether such deposits are eroded, modified, or buried, requires more detailed examination. These missing deposits therefore complicate any estimates of a sediment budget, raising questions about the fate of sedimentary material on Titan.
\newline

\section{Titan's Lacustrine Landscapes}
Like Titan’s fluvial landscapes, the majority of Titan’s lacustrine landscapes are found in the polar regions (Figure \ref{Distribution}; \citet{stofan_lakes_2007,hayes_hydrocarbon_2008,hayes_post-cassini_2018,lopes_global_2020}), presenting a strong reflection of Titan’s current climate \citep{mitchell_climate_2016,faulk_titans_2020}. Unlike Titan's fluvial landscapes, however, the distribution of Titan’s lakes and seas is hemispherically asymmetric, taking up 12\% of the area from 50$^{\circ}$N$-$90$^{\circ}$N but only 0.3\% of the equivalent area in the south \citep{stofan_lakes_2007,hayes_hydrocarbon_2008,hayes_lakes_2016,birch_geomorphologic_2017}. 

Titan’s lacustrine landscapes further cluster within their respective polar regions. Dominating the western hemispheres of both poles are Titan's `Lake Districts' (\citet{mackenzie_case_2019}; Figure \ref{Distribution}). Throughout these regions hundreds of smaller lakes (typically $<$50 kilometers across) in various states of liquid fill are found, all inset into undulating plains thought to be organic-rich \citep{birch_geomorphologic_2017,hayes_lakes_2016}. These lake-cut plains are enclosed in elevated basins, bounded on their perimeters by dissected uplands \citep{moore_landscape_2014,hayes_lakes_2016,hayes_topographic_2017,birch_geomorphologic_2017}. At lower elevations on the eastern hemispheres (Figure \ref{Distribution}) are Titan's broader depressions, both the liquid-filled seas and empty paleoseas \citep{birch_geomorphologic_2017}. These dichotomies suggest either compositional, hydraulic, and/or geophysical controls \citep{birch_geomorphologic_2017,hayes_topographic_2017,hemingway_rigid_2013,hallett_global_2015,faulk_titans_2020,lora_topographic_2022}.
\newline

\subsection{Liquid-Filled North Polar Seas \& Large Lakes}
\begin{figure}[h]
\centering
\includegraphics[width=\textwidth]{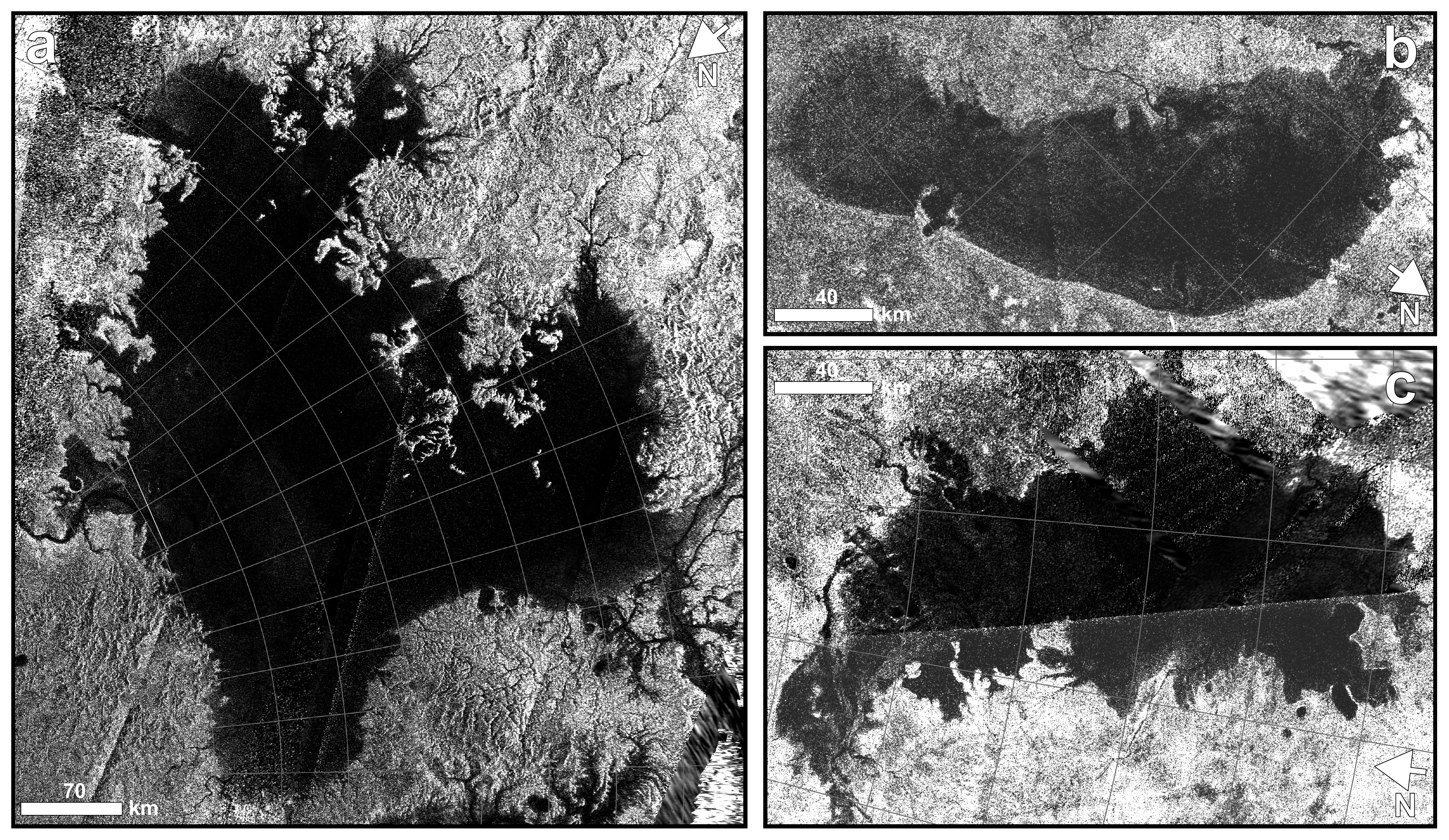}
\caption{\textbf{Diversity of lacustrine landscapes on Titan.} (a) Ligeia Mare at Titan's north, a largely flooded sea, with tentative evidence of wave-cut shorelines, pits, flooded valleys, and no obvious river deltas. (b) Ontario Lacus at Titan's south, with pits along its perimeter and seafloor and, most notably, the only evidence of putative depositional coastal landforms \citep{wall_active_2010}. (c) Jingpo Lacus at Titan's north, a lake comparable in size to Ontario Lacus, typified by dozens of 5$-$10 kilometer pits along its shorelines and seafloor.}
\label{Titan_seas}
\end{figure}

In Titan’s north polar region, broad liquid-filled depressions form three seas - Kraken Mare ($5 \times 10^5$ km$^2$), Ligeia Mare ($1.3 \times 10^5$ km$^2$), and Punga Mare ($0.6 \times 10^5$ km$^2$) - and a handful of the largest lakes including Jingpo Lacus ($0.22 \times 10^5$ km$^2$). These largest features are restricted to the north polar latitudes ($>$50$^{\circ}$N; Figure \ref{Distribution}) and account for $>$80\% of all liquid-filled surfaces by area and $>$95\% of exposed surface liquid by volume \citep{hayes_lakes_2016,birch_geomorphologic_2017}. Where resolution was sufficient to resolve Titan’s coastlines, valley networks are consistently observed to be entering the seas (Figure \ref{Titan_seas}; \citet{miller_fluvial_2021}), appearing flooded based on their crenulated appearance (Figure \ref{Titan_seas}; \citet{stofan_lakes_2007,hayes_transient_2011}). Altimetry measurements over one such network (see above; \citet{poggiali_liquid-filled_2016}) support these interpretations. Kraken Mare, the largest of Titan’s seas and largest enclosed surface liquid body in the Solar System, also appears to consist of a series of broader, topographically controlled basins connected by narrow ($\sim$10 kilometer wide) straits \citep{lorenz_flushing_2014}. What formed these broad basins, whether impacts or tectonism \citep{hemingway_rigid_2013,hallett_global_2015}, and the volume of sediment stored within them both remain unknown. 

Yet the morphology of Titan’s north polar coastlines are not dominated by flooding alone (Figure \ref{Titan_seas}). Consistently observed along both the coastlines and on the seafloor of all the seas and largest lakes are $\sim$5$-$10 kilometer diameter circular depressions (Figure \ref{Titan_seas}). In fact, no large lacustrine body on Titan lacks such features \citep{birch_detection_2022}. Along the coasts of all the large seas, portions of the coastlines that are exposed to a greater fraction of open sea also appear systematically straighter than more protected portions of coast \citep{palermo_coasts_2022}. These observations can be consistent with formation via wind-driven waves, where larger waves intersecting the coastlines along more open coasts are more capable of eroding the shoreline into a smoother shape \citep{huppert_influence_2020,palermo_coasts_2022}. Evidence of waves on Titan’s seas also exists as both the RADAR \citep{hofgartner_transient_2014,hofgartner_titans_2016} and VIMS \citep{barnes_cassinivims_2014} instruments detected surface roughening on Titan’s lakes and seas that could be explained by wind-driven waves. Observations of Titan's coastlines can therefore be leveraged to derive wind speeds necessary to mobilize sediment and/or detach bedrock (\citet{schneck_coasts_2022}, in prep; \citet{palermo_coasts_2022}), highlighting the intimate link between Titan's landscapes and climate. The output of such studies would also serve as an important input dataset for any follow-up mesoscale atmospheric (e.g., \citet{rafkin_air-sea_2020}), climate \citep{lora_topographic_2022} and sea circulation \citep{lorenz_flushing_2014} modeling. Finally, few obvious depositional features are found along the shorelines of any of the north polar lakes and seas, especially river deltas \citep{birch_detection_2022}. On Earth, many rivers terminate in a delta when they intersect standing bodies of liquid \citep{nienhuis_global-scale_2020,nienhuis_deltas_2022}. On Titan, deltas are the exception, with no obvious candidates at Titan’s north \citep{birch_detection_2022}. Why this is the case remains an outstanding puzzle.

Similar to Earth’s oceans, the elevations of Titan’s seas were measured to be consistent with an equipotential surface (Figure \ref{Discovery_timeline}; \citet{hayes_topographic_2017}). Altimetry measurements, from 3 separate flybys (T91, T104, and T108), of the liquid surfaces of Ligeia Mare, Kraken Mare, and Punga Mare fall within 8 meters of each other (relative to Titan’s best-fit geoid), well within expected ephemeris errors in reconstructed spacecraft positioning \citep{hayes_topographic_2017}. For Punga and Kraken Mare, which were observed during the same flyby (T108) and not subject to flyby-to-flyby ephemeris errors, the agreement is better than $1.4$m \citep{hayes_topographic_2017}. These measurements are consistent with morphological observations that the seas are interconnected by valley networks \citep{lorenz_flushing_2014,birch_geomorphologic_2017}. Most importantly, these novel measurements of sea level on an extraterrestrial world suggest that liquid surface elevations of Titan’s largest lakes and seas share a common equipotential surface and, like Earth’s globally averaged sea level, can be used as the reference datum on Titan \citep{hayes_topographic_2017}. As both polar regions are globally the lowest elevations on Titan \citep{corlies_titans_2017,hemingway_rigid_2013,hallett_global_2015}, Titan’s large lakes, seas, and paleoseas should therefore represent the terminal basins for Titan's sediment transport and hydrologic systems, much like the oceans here on Earth.

\begin{figure}[h]
\centering
\includegraphics[width=\textwidth]{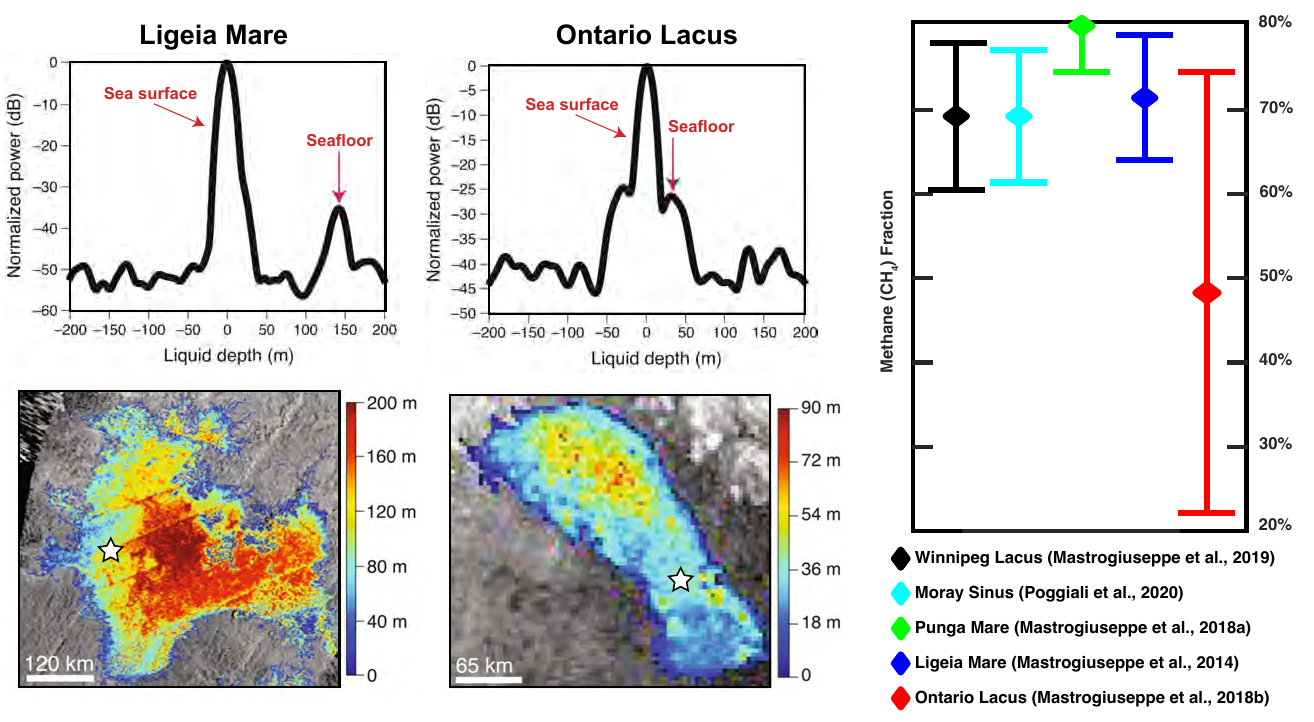}
\caption{\textbf{Bathymetry and composition of Titan's largest lakes and seas.} Top: individual pulses from the T91 and T48 altimetry passes over Ligeia Mare and Ontario Lacus. The first reflection with the highest returned power for each was from the surface of each liquid body. The seafloor reflections were time delayed, and are re-plotted here in terms of depth. The relative power between these two peaks is used to measure the fluid composition. Bottom: Bathymetry maps of Ligeia Mare and Ontario Lacus. Stars denote locations where the above altimetry pulses were acquired. Right: Measured fluid compositions of Titan's lakes and seas from altimetry data. Compositions are plotted as a methane fraction, assuming the nitrogen solubility relation from \citet{malaska_laboratory_2017}}
\label{Titan_bathy}
\end{figure}

The \textit{Cassini} altimeter was also used as a sounder to probe the depth and composition of Titan’s seas (Figure \ref{Discovery_timeline}; \citet{mastrogiuseppe_bathymetry_2014}). During an experiment designed to search for waves on Ligeia Mare in May 2013 (Figure \ref{Discovery_timeline}), the returned altimeter waveforms facilitated an unexpected discovery in the form of secondary returns from the seafloor (Figure \ref{Titan_bathy}).  Along a 300-kilometer track across Ligeia Mare, \citet{mastrogiuseppe_bathymetry_2014} successfully detected multiple subsurface reflections and used the two-way travel time to model depths of up to 160 meters. The relative amplitude of the surface and subsurface returns related to the liquid’s composition, and showed that Titan's methane-ethane-nitrogen fluids are remarkably transparent \citep{mitchell_laboratory_2015} to \textit{Cassini}’s RADAR. This result was unexpected, as it disagreed with experiments conducted by members of the \textit{Cassini} team that found cryogenic methane/ethane to be more absorptive \citep{paillou_microwave_2008}, but did agree with the pre-launch estimates of \citet{thompson_and_squyres_1990}. Immediately following this discovery, a number of \textit{Cassini}’s final Titan flybys (Figure \ref{Discovery_timeline}) were dedicated to acquiring altimetry tracks over the other seas, including central Kraken Mare and its eastern estuary Moray Sinus (T104; August 2014), and both an estuary of Kraken Mare (Baffin Sinus) and Punga Mare (T108; January 2015). While central Kraken Mare proved too deep (or too absorptive) to return a subsurface reflection, the datasets from Moray Sinus, Baffin Sinus, and Punga Mare all revealed similar sub-surface reflections and showed that the interrogated liquids were uniformly transparent and methane-dominated (Figure \ref{Titan_bathy}; \citet{mastrogiuseppe_radar_2016,mastrogiuseppe_bathymetry_2018,mastrogiuseppe_cassini_2018,poggiali_bathymetry_2020}). 

Though only providing snapshots in time, these measurements of uniformly methane-dominated northern seas were unexpected. Because of the lower volatility of ethane, large volumes were expected to be built up over geologic time and deposited in the seas as methane was dissociated in Titan's high atmosphere \citep{lunine_does_1993}. This was consistent with our first, and at the time only,  compositional measurement of Titan's liquids from VIMS in 2007 that showed ethane to be present in Ontario Lacus \citep{brown_identification_2008}. Predictions prior to these flybys also suggested that there should be a latitudinal gradient in the methane-ethane composition that follows Titan's aridity \citep{lorenz_flushing_2014}. Due to higher methane precipitation rates toward Titan's more humid poles (see Chapter 8), poleward seas like Punga Mare were expected to have more methane, while seas like Kraken Mare, which cover lower, more arid latitudes, were expected to be more relatively ethane-enriched \citep{lorenz_flushing_2014}. The general lack of substantial ethane (or any other higher-order hydrocarbon fluids) implied by the lack of microwave absorptivity therefore implies that other mechanisms must be acting to regulate Titan's sea composition. 

One such mechanism to sequester ethane involves losses to a subsurface aquifer \citep{hayes_lakes_2016}. Many lines of evidence point toward the existence of aquifers on Titan (or `alkanofer') including Titan's crater distribution \citep{neish_elevation_2014,wakita_methane-saturated_2022} and lake elevations \citep{hayes_topographic_2017}, measured polar surface temperatures \citep{jennings_surface_2016} and near-surface humidity gradients \citep{lora_near-surface_2017}, observed cloud activity \citep{turtle_titans_2018}, and global circulation climate models \citep{faulk_titans_2020,lora_topographic_2022}. Similarly, circulation within the seas may regulate the composition over shorter timescales. Owing to Titan’s potentially ternary fluid composition \citep{steckloff_stratification_2020}, currents within a Titan sea will be driven by tidal forces, wind stresses, and thermally-driven buoyancy fluxes all expected to be on the order of a few cm/s \citep{tokano_wind-driven_2015,vincent_numerical_2018}. Any single factor therefore cannot be discounted, a complex environment that has yet to be fully explored.  

The RADAR sounder results were also used as ground-truth to calibrate SAR image observations and generate rough bathymetry maps of all of Titan's largest lakes and seas (Figure \ref{Titan_bathy}; \citet{hayes_lakes_2016}). These maps allowed us to estimate the total volume of liquid contained in Titan’s lakes and seas to be $>7 \times 10^4$ km$^3$ \citep{hayes_lakes_2016}, equivalent to a $\sim$1 meter deep global ocean. For comparison, the total volume of the water in Earth’s oceans, if spread equally over the globe, would have a depth of $\sim$2.6 kilometers. This is enormous when compared to the average column depth of total precipitable water in Earth’s atmosphere of only $\sim$22 millimeters. On Titan, the reverse is true. The average column depth of total precipitable methane is $\sim$6 meters \citep{tokano_methane_2006}, or $\sim$6$\times$ greater than the total volume of exposed surface liquid. Therefore, unlike water on Earth, most of the methane in Titan’s current climate is stored in the atmosphere as opposed to surface lakes and seas. As a result, unless there is a substantial subsurface reservoir hidden from remote sensing, Titan’s lakes and seas act more as passive indicators of Titan's climate, as opposed to on Earth where the oceans act as a major buffer that helps regulate global heat transport. Furthermore, these bathymetry maps should prove to be a useful dataset beyond just allowing us to estimate the total liquid volume on Titan, as they can be utilized for any sea circulation, climate, and/or coastal landscape evolution models. 

Finally, despite Titan’s liquid-filled seas being observed multiple times throughout \textit{Cassini}’s 13 years in the Saturn system, and the many mechanisms capable of moving fluids on Titan, only a few changes were ever observed (Figure \ref{Discovery_timeline}). Most obvious were Titan’s `magic islands’, features thought to be either bubbles, floating solids, or waves \citep{hofgartner_transient_2014,hofgartner_titans_2016}, and observed twice in Ligeia Mare over the course of six observations acquired between February 2007 and January 2015 (Figure \ref{Discovery_timeline}; \citet{hofgartner_transient_2014,hofgartner_titans_2016}). The only notable change in shoreline position occurred in the Moray Sinus region, an estuary of Kraken Mare (Figure \ref{Discovery_timeline}; \citet{hayes_transient_2011}). That no further obvious morphological changes took place makes it impossible to discern the rates of shoreline evolution and the impacts of climate variations, or even whether Titan’s shorelines are rocky or sediment-covered. Given \textit{Cassini}’s coarse imaging, however, this is not surprising, as only far finer resolution ($\sim$25 meters) orbital observations of Titan’s coastlines could make the requisite observations.
\newline

\subsection{Ontario Lacus \& Southern Paleoseas}
Few liquid-filled lakes are observed at Titan's south, with 35$\times$ less exposed liquid area than the north (Figure \ref{Distribution}; \citet{birch_morphological_2018,hayes_post-cassini_2018}). Three small lakes are found in a broad plateau between two broader depressions, with only a few other filled lakes ever resolved. Dominating the region is Ontario Lacus, first discovered by ISS \citep{turtle_cassini_2009} and the only lake of significant size ($0.2 \times 10^5$ km$^2$) that was also completely imaged by VIMS, ISS, and SAR (Figure \ref{Titan_seas}) at Titan's south. Like the northern seas and large lakes, Ontario Lacus is characterized by multiple pits \citep{birch_detection_2022}, further stressing the need to understand how such features form. The lake is also distinct in that it is the only large lake/sea to exhibit potential depositional morphologies at \textit{Cassini}'s SAR resolution (Figure \ref{Titan_seas}; \citet{wall_active_2010,hayes_bathymetry_2010,hayes_transient_2011,birch_morphological_2018}). This includes multiple lobate features along the lake’s southern shoreline that are co-located with the terminus of Saraswati Flumen (Figures \ref{Titan_rivers}e \& \ref{Titan_seas}). These features were interpreted as river deltas, with evidence of avulsions given that there are multiple lobes \citep{wall_active_2010}. Unlike Mars, where the crater walls act as a boundary condition that prevents the formation of multiple lobes (e.g., Jezero Crater; \citet{goudge_sedimentological_2017}) and massive breaching floods are a major geomorphic process \citep{goudge_importance_2021}, delta formation on Titan might be more like Earth, where rivers can freely avulse along a low-sloping shoreline. 

Across the rest of the southern shoreline of Ontario Lacus are other putative deposits, with the overall curvature of the shoreline perhaps indicative of a broader alluvial plain. Because sediment transport calculations show that the deposits can readily form in Titan’s current climate \citep{birch_geometry_2022}, their rather diminutive size may be related to the stability of the shoreline position. One such possibility is that only recently, but not necessarily for the first time, was the shoreline stationary at its current position, with the rest of the sediment upslope being re-worked downstream into the new deltas. This is consistent with estimates of long-term climate variations on Titan, where fluids are expected to be preferentially accumulating at Titan’s north, at the expense of the south (\citet{aharonson_asymmetric_2009,hayes_post-cassini_2018,lora_topographic_2022}, see Chapter 8).  

While Ontario’s western shoreline is complex, its eastern shoreline is remarkably smooth (Figure \ref{Titan_seas}; \citet{wall_active_2010}), and remains the clearest evidence for a wave-modified sediment-covered shoreline on Titan (i.e., a beach). Supporting this interpretation are that: (1) the lake sits at the deepest portion of a broader sedimentary basin suggesting the shorelines are sediment covered, (2) wind directions are consistent with the expected along-shore transport direction \citep{wall_active_2010,schneck_coasts_2022}, and (3) small scale morphologies including putative barrier islands \citep{birch_detection_2022} and a build-up of sediment on either side of small lake that cuts the shoreline \citep{birch_detection_2022}. Detailed sediment transport modeling therefore promises to reveal more about the nature of Titan’s current climate. 

Yet at all remote sensing wavelengths, albeit in the potentially less windy 'off-season' \citep{hayes_wind_2013}, Ontario Lacus was observed to be remarkably flat (Figure \ref{Discovery_timeline}; \citet{turtle_cassini_2009,brown_identification_2008,barnes_shoreline_2009,hayes_bathymetry_2010,wye_smoothness_2009}, indicating that winds were insufficient to generate detectable waves, that threshold wind speeds occurred so infrequently that \textit{Cassini} did not observe them, and/or that process(es) may be acting to suppress activity \citep{hayes_wind_2013,cordier_floatability_2019,yu_single_2020}. Titan's northern seas were also observed to be similarly flat and created a significant conundrum for many years. Only later in the mission did sea surface activity finally pick up at Titan's north (Figure \ref{Discovery_timeline}; \citet{barnes_cassinivims_2014,hofgartner_transient_2014,hofgartner_titans_2016}). Such activity, if comprising waves, may be correlated to Titan approaching its northern summer and the increased wind speeds predicted in that season \citep{hayes_wind_2013}.  

At infrared wavelengths, VIMS showed that Ontario Lacus possesses a series of expanding bathtub rings that are bright at 5 $\mu$m and have been interpreted as organic evaporites deposited along paleo shorelines as liquid evaporated and the shoreline receded \citep{barnes_shoreline_2009}. \citet{turtle_shoreline_2011} and \citet{hayes_transient_2011} also argue for shoreline recession at Ontario Lacus between June 2005 and March 2009 (Figure \ref{Discovery_timeline}). The detection of ethane on the lake surface by VIMS \citep{brown_identification_2008} and the relative enrichment in ethane through the bulk fluid column (Figure \ref{Titan_bathy}; \citet{mastrogiuseppe_bathymetry_2018}) further support that evaporation rates may have outpaced runoff outputs, leading to shoreline recession. Note, though, that this shoreline recession is not definitive given serious resolution limitations of the VIMS, ISS, and SAR datasets \citep{cornet_edge_2012}. If Ontario’s shoreline did move, however, and since we know the northern shorelines did not, this discrepancy may be a reflection of the surrounding topography and lake bathymetry, with shallower slopes around Ontario Lacus allowing for more drastic, and therefore more observable changes.

More broadly, the south polar region contains four SAR-dark depositional basins identified as potential paleoseas from a past climate (Figure \ref{Distribution}; \citet{birch_morphological_2018}). These basins are similar in area to the northern liquid-filled seas, and their boundaries appear consistent with a common elevation \citep{birch_morphological_2018,corlies_titans_2017}. These basins lack any 5 $\mu$m-bright evaporitic material \citep{mackenzie_compositional_2016}, however, suggesting burial by sediment transported by winds and/or runoff. The two largest lakes, Tsomgo Lacus and Ontario Lacus, are found at the center, and presumably deepest portions, of the Romo and Ontario basins, respectively. Volumetrically, all four basins are $\sim$1.2$\times$ larger than the northern seas; however, they appear far less interconnected than the northern seas. Where the northern seas are all visibly connected by multi-kilometer wide valleys, at the south, only the Romo and Ontario basins have even the potential to connect if they were filled \citep{birch_morphological_2018}. The boundaries of each of these basins also show evidence of being highly dissected by small-scale valley networks that terminated at the putative shoreline, consistent with geologically recent exposures \citep{hayes_lakes_2016}. In at least one case, incision at a basin boundary shows incision of exposed terrain that appears to propagate away from the shoreline, into pre-existing and less dissected uplands (Figure \ref{Titan_rivers}d; \citet{birch_morphological_2018}). 
\newline

\subsection{Filled \& Empty Lakes}
\begin{figure}[h]
\centering
\includegraphics[width=\textwidth]{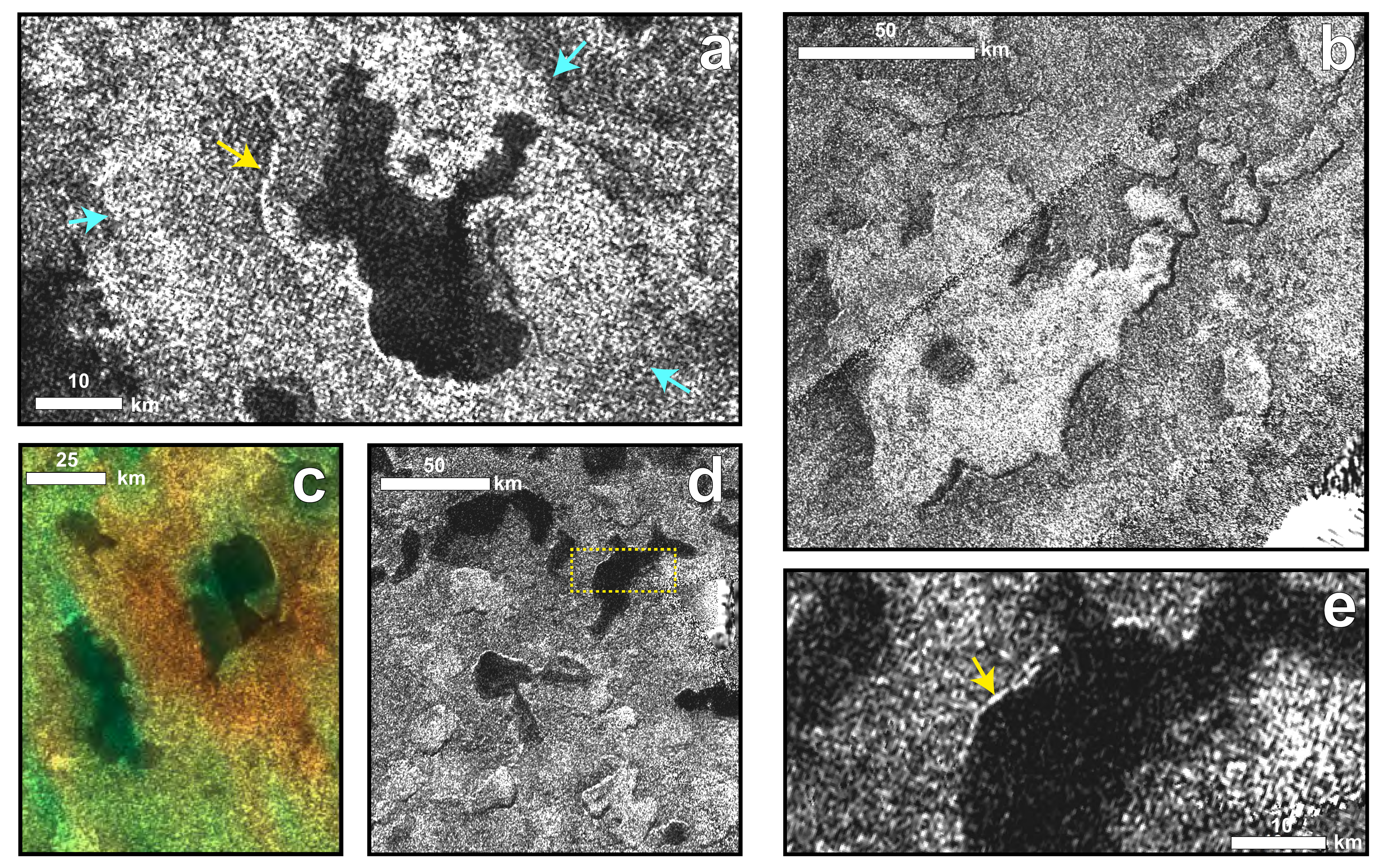}
\caption{\textbf{Titan's Filled and Empty Lakes.} (a) Viedma Lacus at Titan's north, one of the few lakes with both raised rims (yellow arrow) and a surrounding rampart (cyan arrows). (b) Unnamed empty lake at Titan's south exhibiting many of the classic morphologies of Titan's sharp-edged depressions (SEDs) including flat floors, steep boundaries, and evidence of agglomeration. (c) False-color VIMS image overlain on SAR data, re-created from \citet{barnes_organic_2011}, showing 5 $\mu$m-bright material (reddish-brown color) around a north polar lake. (d) Dozens of lakes at Titan's north, with both filled and empty lakes in close proximity to each other. Many of these lakes have elevated rims \citep{birch_raised_2019}, classifying them as SEDs. (e) An expanded view into the highlighted region in panel c, centered on Winnipeg Lacus to highlight the raised rim that reaches $>$100 meters above the surrounding terrains \citep{poggiali_high-resolution_2019}}.
\label{Titan_lakes}
\end{figure}

In total, \textit{Cassini} discovered $\sim$600 filled lakes/seas and $\sim$300 empty-lake depressions \citep{hayes_hydrocarbon_2008,birch_geomorphologic_2017}. The size distribution of these features is log-normal with a median diameter of 77$\pm$20 kilometers \citep{hayes_lakes_2016}. Titan’s filled lakes make up an area of $\sim 2 \times 10^5$ km$^2$ while empty lakes encompass $\sim 1.4 \times 10^5$ km$^2$. At the north, most of Titan’s lakes are filled, though filled and empty ones are often in close proximity (Figure \ref{Titan_lakes}c; \citet{birch_geomorphologic_2017}), suggesting that Titan’s landscape once held larger liquid volumes \citep{birch_geomorphologic_2017}. Meanwhile at the south, almost all of the lakes are currently empty (Figure \ref{Distribution}; \citet{birch_geomorphologic_2017}). All told, Titan’s filled and empty lacustrine landscapes cover $\sim$1.1\% of Titan's global surface area \citep{hayes_lakes_2016}. Comparatively, the Earth has between 50 and 300 million active lakes and ponds that cover $\sim$2.7\% of Earth’s surface \citep{downing_global_2006,mcdonald_regional_2012}.

While Titan’s larger lakes and seas appear to have formed through inundation of a pre-existing landscape, the smaller lakes, Titan’s sharp-edged depressions (SEDs), appear to have originated through entirely different processes \citep{hayes_topographic_2017}. SEDs are found throughout Titan’s north and south polar regions though they largely cluster on the western hemispheres (Figure \ref{Distribution}; \citet{birch_geomorphologic_2017}). They do not exhibit any obvious planform orientations or spatial patterns, such as forming along lines or collecting in definitive clusters \citep{hayes_hydrocarbon_2008,dhingra_using_2019}. They also appear to be topographically closed, with no evidence for inflow or outflow networks at the resolution of \textit{Cassini} images (Figure \ref{Titan_lakes}). Empty SEDs have an average depth, relative to surrounding terrain, of 200$\pm$100 meters and show no significant correlation between depth and plan-view diameter \citep{hayes_topographic_2017}. Such depths are significant, rivaling those measured for the seas \citep{hayes_lakes_2016}. 

The majority of both filled and empty SEDs also exhibit 100-meter high, kilometer wide, raised rims that rise above the lake floors and surrounding terrain (Figure \ref{Titan_lakes}d; \citet{birch_raised_2019}). These rims are observed to be just a few kilometers wide in both \textit{Cassini} SAR images (e.g., Figure \ref{Titan_lakes}e) and altimetry data \citep{michaelides_constraining_2016,birch_raised_2019,poggiali_high-resolution_2019}, and present a significant challenge to formation models for Titan's SEDs. Especially problematic is that SED planform shapes suggest that they grow radially via scarp retreat \citep{hayes_topographic_2017}, appearing as agglomerations of multiple smaller features (Figure \ref{Titan_lakes}a,b,d). If true, this would require destruction and subsequent regeneration of the raised rim as the SED expands. Alternatively, if the SEDs grew laterally in a past, more favorable climate, then the rims may represent a late stage addition, perhaps via chemical interactions along the boundary that indurate their perimeters. Such a rim would therefore represent a type of inverted relief. Regardless, most formation models for SEDs invoke an unexplained shift in Titan’s climate (e.g., \citet{mitri_possible_2019}), with none yet able to satisfactorily explain the many, often conflicting, constraints provided by analyses of \textit{Cassini} data \citep{hayes_topographic_2017}.

A small number of the SEDs ($<$10) are also surrounded by elevated annuli of SAR-bright material termed ramparts (Figure \ref{Titan_lakes}a; \citet{solomonidou_spectral_2020}). Unlike the rims, these ramparts are 10’s of kilometers wide, but are found only on select SEDs, all clustered on top of a single, elevated plateau in Titan’s northern lake district \citep{birch_geomorphologic_2017}. These features may point to an earlier stage of SED evolution \citep{solomonidou_spectral_2020} or they may form by an entirely different, more explosive, process like cryovolcanism \citep{wood_morphologic_2020} or nitrogen maar-like explosions \citep{mitri_possible_2019}. For clarity, these ramparts are extremely rare and are far wider compared to the raised rims (Figure \ref{Titan_lakes}a) that are found around the majority of Titan’s small lakes \citep{birch_raised_2019}. 

As it did for the large seas, \textit{Cassini} also measured the liquid elevation of eleven small lakes scattered throughout the north polar terrain. The liquid elevation of these lakes was found to be hundreds of meters above the elevation of the seas, suggesting that they reside in isolated or perched drainage basins that may or may not be hydrologically connected to the seas \citep{hayes_topographic_2017,poggiali_bathymetry_2020}. Further, the floor elevations of empty SEDs appear to sit above the elevation of filled lakes or seas, measured via RADAR altimetry, that are in the same drainage basins \citep{hayes_topographic_2017}. This suggests that the empty SEDs whose floors go below filled lakes or seas in the same drainage basin become filled themselves, hinting at subsurface connectivity and communication through an aquifer.

\textit{Cassini} also constrained the liquid composition of two lakes during the T126 flyby (Figure \ref{Discovery_timeline}). One lake in particular, Winnipeg Lacus (Figure \ref{Titan_lakes}c/d), was large enough to provide eight independent lake floor returns. The deepest of these detections was $>$100 meters, suggesting a low loss tangent and methane-dominated composition similar to the seas (Figure \ref{Titan_bathy}; \citet{mastrogiuseppe_deep_2019}). Methane-dominated liquid compositions was a surprising result for the small lakes, as morphological evidence points to formation models that involve dissolution-based processes \citep{hayes_topographic_2017}. If dissolution were involved in the formation of the SEDs, it is reasonable to expect that the liquid would retain dissolved higher-order and/or potentially polar organics with significantly higher loss tangents as compared to liquid methane. The fact that the microwave transparency of the small lakes is similar to the larger seas suggest that either: (1) the loss tangent is not affected by the presence of dissolved loads or solutes, (2) that the liquid in the lakes has been flushed by precipitation and/or runoff-derived fluid that has not yet been saturated in dissolved components, or (3) Titan’s SEDs are no longer dissolving (or dissolution never occurred) along their perimeters, and are merely basins that passively fill if they intersect the subsurface aquifer. The 5 $\mu$m-bright deposits interpreted as evaporites that have been found in the floors of dry SEDs and ringing some liquid-filled SEDs (Figure \ref{Titan_lakes}c; \citet{barnes_organic_2011,mackenzie_evidence_2014,mackenzie_compositional_2016}), however, suggest that dissolved components are at least intermittently present near Titan’s SEDs. Therefore, as is the case for many open science questions regarding Titan, for every answer that the \textit{Cassini} dataset provides, several new and exciting questions emerge for future studies and missions to address!

Though the structural boundaries of these lakes were not observed to change, multiple observations of fluids migrating near these lakes were made (Figure \ref{Discovery_timeline}). \citet{turtle_cassini_2009} and \citet{turtle_rapid_2011} discuss dark features that appeared in a topographic depression in the vicinity of Arrakis Planitia between ISS observations acquired in July 2004 and June 2005 (Figure \ref{Discovery_timeline}), shortly following southern summer solstice and bounding a large-scale south polar cloud outburst observed October 2004 \citep{schaller_storms_2009}. Such features are most consistent with a vast surface wetted by rainfall \citep{turtle_titans_2018}. \citet{hayes_transient_2011} discuss repeated RADAR passes of the South acquired in 2007 and 2008/2009 that show that liquids within the bottom of partially filled lakes seem to disappear between subsequent SAR observations (Figure \ref{Discovery_timeline}). Similarly, \citet{mackenzie_case_2019} described a collection of northern lakes that disappeared between SAR and VIMS observations in 2006 and 2013, respectively (Figure \ref{Discovery_timeline}). In both cases, the observed difference in SAR backscatter and/or infrared albedo cannot be explained by geometric effects and suggests that, between the observations, liquid either infiltrated into the ground, evaporated, or did both \citep{mackenzie_case_2019}. These changes suggest that the fluids in Titan’s small lakes are directly influenced both by flow of fluid into a subsurface aquifer and by evaporation/precipitation from the atmosphere. With a dedicated observing platform in orbit around Titan, we could better constrain the timescales of these interactions.
\newline

\subsection{Low-Latitude Lacustrine Features}
While the existence of stable (i.e., long-lived) equatorial lakes has been proposed based on Earth-based radar observations (e.g., \citet{campbell_radar_2003}), \textit{Cassini} VIMS observations of kilometer-scale low albedo features \citep{griffith_possible_2012,vixie_possible_2015}, and theory \citep{tokano_stable_2020}, no liquid-filled features have been confirmed by finer resolution SAR images or nadir-pointing RADAR datasets \citep{hofgartner_root_2020}. As a result, confirmed liquid-filled lakes remain restricted to polar latitudes (50$^{\circ}$N$-$90$^{\circ}$N and 50$^{\circ}$S$-$90$^{\circ}$S, respectively), where colder temperatures \citep{jennings_surface_2016,jennings_titan_2019} allow for persistent lakes and seas in Titan’s current climate. 

Yet there is evidence of past low latitude lakes and seas, perhaps stable under a previous climate regime. Buried and/or highly eroded SEDs appear to extend down to $\sim$40$^{\circ}$ latitudes in the Soi Crater \citep{solomonidou_soi_2022} and South Belet \citep{schoenfeld_geomorphological_2021} regions. Most striking, the southern tropical regions of Tui and Hotei Regios (Figure \ref{Distribution}) are in regional lows \citep{corlies_titans_2017}, with valley networks draining the surrounding mountainous terrains into fields of SAR-bright, lobate depressions that are morphologically similar to polar lakes \citep{moore_landscape_2014}. These regions also correspond to the largest non-polar exposures of 5 $\mu$m-bright evaporite deposits \citep{mackenzie_compositional_2016}, and hosted near-specular reflections from the Arecibo telescope \citep{hofgartner_root_2020} like the northern empty lakes \citep{michaelides_constraining_2016}. All of these observations strongly suggest that such features are dry lake beds, with alternative hypotheses that Tui Regio and Hotei Regio represent cryovolcanic deposits \citep{soderblom_geology_2009} less likely than was thought earlier in \textit{Cassini}'s mission. As such, unless the empty lakes within Tui and Hotei Regios can form rapidly and without stable surface liquids, their presence suggests that Titan’s low latitudes may have been home to surface liquids, at least transiently, in the recent past. Accordingly, Titan’s climate may have experienced significant drying over its lifetime due to global methane losses \citep{hayes_lakes_2016}, something that follow-up climate models should explore. 
\newline

\section{Outstanding Open Questions Post-\textit{Cassini}}
We are in an exciting era, when the \textit{Cassini} data has allowed us to ask questions we never thought possible just two decades prior. Simultaneously, we eagerly await yet more questions we will soon be able to ask from \textit{Dragonfly}. Though much \textit{Cassini} data analysis remains to be done, many of the largest questions (see below) now also require new quantitative investigations, employing the tools and methodologies that have been used on Earth and Mars for decades. Two key themes that we will explore here are that Titan presents opportunities to learn about the basic processes that govern planetary hydrologic cycles and about how planetary climates are maintained through time.

Of particular interest is that Titan's crustal materials, its `rocks,' are unlike anything we experience on terrestrial rocky planets. Complex organic molecules are synthesized throughout Titan's atmosphere and fall to the surface, where they can interact with the water ice that comprises Titan's lithosphere. Its fluids are also cryogenic methane and ethane, in different proportions and with different properties depending on local-scale processes \citep{steckloff_stratification_2020}. Interactions between these materials, though yet unknown, remain astrobiologically intriguing as it is thought that the chemistry occurring on Titan's surface today may not be dissimilar from the pre-biotic chemistry that took place on Earth's surface billions of years ago \citep{trainer_organic_2006}. For this very reason, the \textit{Dragonfly} mission was selected in 2019 to explore Titan's surface, with many of the mission's primary objectives to explore this complex chemistry in situ \citep{barnes_science_2021}. 

Beyond the astrobiological importance, however, is the opportunity to explore how a hydrologic system operates with entirely new materials, as we have only ever explored hydrologic systems on rocky worlds. Further, the only other active system is on Earth where vegetation, ices, and humans play important, complicating roles. Titan therefore offers the chance to understand how the basic processes that govern an active hydrologic system are influenced by truly alien materials, all of which are occurring within what is, perhaps, a simpler environment than here on Earth.    

Planetary exploration has also shown that stable hydrologic systems are the exception, not the norm. Venus and Mars both may have once had environments like Earth, but they are since long lost and only limited geologic evidence remains. Titan is the only other location in the solar system where it is possible to understand how an active planetary climate and hydrologic cycle are maintained through time (also see Chapter 8). This is especially important as we are now capable of exploring exoplanetary atmospheres and climates, models for which are largely based on processes relevant to the present-day Earth \citep{kasting_habitable_1993}. As the worlds likely to be explored by next-generation telescopes have similar, or even greater, diversity than in our own solar system, using Titan as a second data point to study planetary climate-surface dynamics will be critical. 

We therefore want to understand the longevity of methane in Titan's atmosphere, a question that remains one of the most outstanding post-\textit{Cassini} (\citet{nixon_titans_2018}, and see Chapter 8). Are there processes or feedbacks capable of stabilizing Titan's current, methane-based planetary climate, like the mechanisms that have been hypothesized here on Earth \citep{berner_carbonate-silicate_1983}? Has Titan had multiple, intermittent different climates? Or is its present atmospheric methane just a late, transient addition (e.g., \citep{charnay_titans_2014})? Just as Earth's fluvial and coastal landscapes retain important information about the effects of both the current climate, and longer term climate variations, so too should Titan's. Yet few quantitative studies have yet been carried out that focus on Titan’s rivers, hillslopes, and coasts. As a result, our knowledge of what Titan’s landscapes can reveal about Titan’s active climate and sedimentary processes remains only in its infancy.    

Much of this deficiency is owed to the fact that much of the data typically available to terrestrial and martian geologists, especially topographic data and field observations, are coarse and/or missing for Titan. To remedy this, and to address many of the above broad science themes, new methods that test and ground quantitative models must be developed in tandem (e.g., \citet{black_estimating_2012,tewelde_estimates_2013,black_global_2017,birch_geometry_2022,palermo_coasts_2022}). Titan’s active environment therefore provides an exciting playground for geomorphologists, oceanographers, and climate scientists to explore over the coming decades, with its active hydrologic cycle offering the greatest opportunity to better understand planetary climates and hydrologic cycles more generally. Though the data will remain limited even with \textit{Dragonfly}, many of the outstanding science questions listed below could be addressed by a Titan orbiter \citep{mackenzie_titan_2021}, one capable of resolving the myriad expected landscapes and searching for changes on the surface as spacecraft orbiting the Earth do regularly (e.g., Landsat). That such a mission was prioritized in the 2023$-$2032 Planetary Science Decadal Survey, Origins, Worlds and Life \citep{OWL_2023}, gives hope that such observations will be made in the coming decades. Below we list specific research directions that could shed light on Titan's hydrologic cycle and the mechanisms that may maintain its planetary climate through time.
\newline 

\textit{1. \underline{How does bedrock break down on Titan's hillslopes?}} \newline
The rates and mechanisms through which rivers are incising on Titan, the volumes of sediment they convey (if any), and the degree to which Titan’s topography has been eroded, all relate to Titan's surface material properties. All remain understudied, as we have little information about how sediment is sourced on Titan’s hillslopes, the rates at which sediment is delivered to river channels, and whether such rivers even have the necessary tools to incise mechanically (e.g., \citet{sklar_mechanistic_2004}). Despite being acknowledged as a major question a decade ago \citep{griffith_titans_2014}, little progress has been made on the topic of Titan’s hillslopes, presenting a major knowledge gap since hillslopes on Titan likely occupy the vast majority of the landscape, as they do on Earth \citep{perron_formation_2009}. Titan's hillslopes are also expected to be a major source of sediment, and by area, are where most of the astrobiologically-relevant chemical evolution of Titan's crust and surface materials will take place. Understanding hillslope processes on Titan is therefore essential.

On Earth transportable sediment is largely generated via weathering processes on hillslopes. Though weathering processes are often tied to biology \citep{richardson_influences_2019}, many also result from thermophysical, chemical, or physical mechanisms \citep{heimsath_soil_1997,heimsath_soil_2000,wilcock_geomorphic_2013,dixon_climate-driven_2009,hall_thermal_2014}. Additional planetary-specific weathering mechanisms include impact gardening or volatile ice sublimation, both thought to be important for generating sediment on airless bodies \citep{howard_sublimation-driven_2008}. 

On Titan, if the bedrock composition of hummocks and crater rims is water ice or ice clathrate, it is not clear if any known mechanisms are capable of weathering and producing appreciable volumes of sediment. Because Titan's rivers have likely incised, forcing the hillslopes to respond, and because producing significant quantities of sediment may therefore be difficult, Titan’s hillslopes may be bare bedrock, with sediment later delivered to channels only along steep slopes via landsliding above some failure criterion \citep{griffith_titans_2014}. Any rounded, sediment-covered hillslopes detected by \textit{Dragonfly} along the Selk crater rim would therefore be both surprising and of enormous interest. Such observations would greatly help in understanding the make-up of Titan’s bedrock material, the mechanisms and rates at which it is weathered to produce sediment, the processes (e.g., rain splash, 'clay-like' swelling, etc.) responsible for transporting sediment down slope, and the rate at which channels are incising along the hillslope boundary.
\newline

\textit{2. \underline{How much of Titan's topography have rivers exhumed?}}  \newline
In tandem with understanding hillslope weathering mechanisms, understanding the degree to which rivers have incised on Titan should also allow for assessments of whether there is ongoing tectonism \citep{black_global_2017}, helping address fundamental questions related to Titan’s crustal composition, interior structure and thermal evolution. The production of new relief on Earth maintains slopes steep enough to transport sediment. In most cases, the original topographic surface is lost as river networks completely modify the surface. Without new relief generation on Titan, either through impacts or tectonism, its rivers may be armored \citep{howard_formation_2016,birch_alluvial_2016}, and its current relief evolution set to `pause’ unless significant climate variations provide additional river discharge. Alternatively, Titan's topography may continue to slowly erode until it reaches a base level, where little relief is left \citep{hack_1975}. In both cases, its present drainage networks may relate to the original topographic surface, perhaps like that of the impact- and tectonic-dominated surface of Ganymede \citep{neish_fluvial_2016}. 

\textit{Cassini} showed that eroding river networks may exist on Titan today. Vid Flumina is incised 100’s of meters into the surrounding terrain \citep{poggiali_liquid-filled_2016}, and the dendritic networks observed by \textit{Huygens} are incised a few tens of meters and slope quite steeply towards the south ($\sim$5\%; \citet{daudon_new_2020}). On Earth, large slopes on the order of 5\% suggest a detachment limited network \citep{birch_geometry_2022}, wherein bedrock material is present along the bed and/or banks. Topographic data of any other valley network on Titan are insufficient \citep{corlies_titans_2017} to make similar conclusions, both due to coarse vertical resolution, and their inability to resolve the channels themselves. New topographic data of Titan’s surface are therefore critical \citep{mackenzie_titan_2021} to quantitatively confirm the presence of incised river channels, and then understand their dynamics. Incising channels along the steeper portions of the rim of Selk crater imaged by \textit{Dragonfly} during its search for frozen impact melt ponds \citep{barnes_science_2021} would also aid greatly in our study of such rivers and allow for inferences about Titan's material properties and tectonic history.  
\newline

\textit{3. \underline{What surface processes help regulate Titan's long-term climate?}} \newline
The degree to which Titan’s topography has been exhumed, and the rate over which it occurs, are also directly related to the timescale over which Titan’s methane hydrologic cycle, and therefore the methane in the atmosphere, has persisted. Evidence from the shape of drainage basins \citep{miller_fluvial_2021} and landscape evolution models \citep{black_estimating_2012,tewelde_estimates_2013} suggest that limited erosional exhumation has occurred near Titan’s poles. Yet Titan's overall surface is quite young \citep{neish_titans_2012}, suggesting that its methane may have persisted for geologic timescales. 

These contrasting observations suggest one of three scenarios (see also Chapter 8). First, Titan's atmospheric methane may be ancient, but erosion rates are very slow. Second, Titan's atmospheric methane may be recent, yet most of the evolution of its surface may have occurred under a previous, nitrogen-only atmosphere \citep{charnay_titans_2014}. Third, Titan's topography may be relatively new, perhaps due to dynamics in the broader Saturn system \citep{wisdom_2022}, such that it has not had sufficient time to be significantly eroded. 

All three scenarios relate to the age of Titan's atmospheric methane and whether it is ancient, a recent addition, or not even the primary fluid responsible for most of its surface evolution. Without relying on re-supply from Titan's interior, one method that may regulate Titan's atmospheric methane and its current climate would be the chemical weathering of minerals in its near-surface crust (e.g., co-crystals, \citet{cable_co-crystal_2019,cable_titan_2021,maynard-casely_co-crystal_2016,czaplinski_experimental_2020}). On both Earth, and likely ancient Mars, chemical weathering of bedrock consumes atmospheric CO$_2$ and acts as a planetary thermostat, stabilizing the global climate over geologic timescales \citep{berner_carbonate-silicate_1983,kasting_habitable_1993}. If indeed such a mechanism exists on Titan, its atmospheric methane may be ancient, much like Earth's water,  offering another complication to consider as we continue to explore exoplanetary atmospheres \citep{kasting_habitable_1993}. To explain the resultant low erosion rates, laboratory and in situ \textit{Dragonfly} measurements of the physical properties of Titan materials would also be of high value. Due to Titan's fluid properties and the low submerged weight of potential sediments, additional yet unexplored dynamics may also occur within a Titan river that buffer erosion rates. Laboratory and numerical investigations of such dynamics, and how Titan analog materials interact with cryogenic fluids, are therefore necessary.   

If Titan's atmospheric methane is young, then numerous new questions arise. Would such a climate shift be consistent with other landscapes observed across Titan? Why have Titan's polar regions exhibited little exhumation while its equatorial and mid-latitudes are buried under dune sands? Are new observations required to time the potential shift in Titan's climate, and if so, what landscapes would be most diagnostic? How would sediment transport and landscape evolution proceed in a nitrogen-only hydrologic cycle? For all these questions, and others not listed here, new numerical, laboratory, and analytic studies are clearly needed to investigate the impacts on the long-term climate and landscape evolution. 

Advancements in our understanding of the rates and magnitude over which Titan’s surface topography is eroding would be among the most important studies in Titan science as a whole, as the implications touch on many of the most outstanding questions post-\textit{Cassini} \citep{nixon_titans_2018}. Necessary for advancement are both theoretical and numerical studies of the rate over which Titan’s topography has formed, with the aim of deriving meaningful timescales. Additional topographic data would also be of enormous value to addressing this question \citep{mackenzie_titan_2021}, enabling relative estimates of exhumation rates between different terrains. \textit{Dragonfly}’s sampling and compositional analyses of Titan’s surface \citep{barnes_science_2021}, in particular the bedrock material, would enable yet more quantitative investigations into the rates of weathering and erosional exhumation. Finally, a global imaging and topographic dataset \citep{mackenzie_titan_2021} at resolutions comparable to \textit{Huygens} and capable of detecting changes over multiple years would allow for the full toolkit of quantitative geomorphology to be employed. With such a dataset, the active coupling between erosion, climate, and tectonism could at long last be explored on a planet other than the Earth.  
\newline

\textit{4. \underline{How important is chemical erosion on Titan relative to mechanical erosion?}}\newline 
Given its analogy to other icy moons, much of the Titan literature early in \textit{Cassini}’s mission assumed the bedrock was largely water ice (e.g., \citet{griffith_titans_2014}, and references therein). Because water ice is insoluble in liquid hydrocarbons over any reasonable timescale \citep{lorenz_erosion_1996}, chemical erosion was not thought to be important. Titan's river valleys also form continuous, organized networks that extend for 100's of kilometers (Figure \ref{Titan_rivers}). That no known networks on Earth form vast, continuous networks via dissolution \citep{griffith_titans_2014} further suggests that mechanical erosion into water ice is significant across large swaths of Titan.

However, dissolvable species are now thought to be present on Titan’s surface, at least regionally. This has been suggested based on photochemical models (e.g., \citet{krasnopolsky_photochemical_2010}), and evidenced by detections of dissolvable surface materials \citep{singh_acetylene_2016} and in the form of evaporite deposits around Titan’s empty lakes (Figure \ref{Titan_lakes}c; \citet{barnes_organic_2011,mackenzie_evidence_2014,mackenzie_compositional_2016,cordier_chemical_2013,cordier_structure_2016,cable_co-crystal_2019,czaplinski_experimental_2019}. Morphologically, two terrains also signal that dissolution erosion, at least regionally, may be important: Titan’s sharp-edged depressions \citep{hayes_topographic_2017,birch_geomorphologic_2017} and the labyrinth terrains \citep{malaska_labyrinth_2020}. Though not a surefire sign of chemical erosion given the relatively low resolution of \textit{Cassini}’s remote sensing datasets, these two terrains most suggest that chemical erosion may occur, at least regionally, on Titan.  
 
Follow-up studies using both \textit{Cassini} data and theory should investigate to what degree dissolvable species make up Titan’s bedrock, how that varies across Titan, and how such materials are incorporated into Titan’s rock cycle. Modeling of detachment limited erosion of channels on Titan \citep{neish_fluvial_2016,howard_formation_2016} could quantify the role that water-ice rich terrains, globally distributed as they are \citep{lopes_global_2020}, hold in the long term evolution of Titan’s topography and sediment budget (see above), as it did for Mars \citep{howard_simulating_2007}. Theoretical and laboratory studies of how a landscape evolves when both mechanical and chemical erosion are influential would have similar value, allowing for definitions of when a landscape transitions from one erosional type to another. Quantitative comparisons, though difficult, could then be made to Titan's existing landscapes. Such modelling work must also be supported by laboratory measurements so that meaningful rates could be constrained. Fortunately, laboratory investigations on the chemical \citep{cable_co-crystal_2019,cable_titan_2021,maynard-casely_co-crystal_2016,czaplinski_experimental_2020,czaplinski_experimental_2019,hanley_titan_2017,farnsworth_nitrogen_2019} and mechanical \citep{collins_relative_2005,litwin_influence_2012,maue_rapid_2022,yu_single_2020} behavior of various Titan fluids and materials are making significant progress. Significant work remains, however, and is of high priority, in particular measurements of the solubility, strength, and density of such materials.
\newline

\textit{5. \underline{What can Titan's rivers reveal about its materials and climate?}}\newline
Whether a river's channel bed and banks are sediment covered (termed `alluvial') or have exposed bedrock is important for setting the river's geometry and the rates at which it modifies the landscape through erosion and deposition. Both a river's geometry and its deposits are observable in remote sensing data and diagnostic of the planetary climate. It is therefore necessary to discern whether a channel on Titan is erosional in nature, or whether there are self-formed alluvial channels flowing through their own sediment. 

In the highly unlikely scenario where finer resolution image and topographic data showed that Titan lacks alluvial rivers entirely, then our understanding of how alluvial channels form and transport sediment would require serious revisions. Therefore, assuming for now that some of Titan’s rivers are alluvial, their geometry could be used to infer the rates of flow and sediment within the channels \citep{birch_geometry_2022}, both of which are reflective of Titan’s active climate. Though complicated somewhat by Titan’s potentially unique ternary fluid composition \citep{steckloff_stratification_2020}, and its infrequent but large magnitude, long duration storms \citep{faulk_regional_2017,battalio_global_2021,battalio_interaction_2022}, terrestrial relations that predict the flow and sediment transport rates in rivers should also be broadly applicable to Titan. So long as implicit empiricisms are handled carefully, similar such relations promise to provide new insight into the materials that form on Titan's surface and the nature of its active climate. 
 
Whether alluvial rivers on Titan form a single channel (or `thread') rather than multiple braided threads would also inform as to the nature of the sediments that such rivers traverse. A single-threaded alluvial river would suggest that the sediment must have some cohesion along the channel banks \citep{howard_how_2009,braudrick_experimental_2009}. Such rivers may exist at Titan’s south (Figure \ref{Titan_rivers}c; \citet{malaska_high-volume_2011}), though we have little idea if we are observing the channel banks or valley walls. If indeed such rivers are confirmed to exist, laboratory investigations would still be required to understand whether confinement is due to the flocculation of Titan materials, or other organic solids, tholin-like aerosols, and co-crystals \citep{cable_co-crystal_2019,yu_single_2020,cable_titan_2021}. A braided or multi-threaded channel meanwhile requires minimal cohesion (for equal sediment loads), and would form far broader channels. Without vegetation \citep{dong_roles_2019,zeichner_early_2021} or significant quantities of mud \citep{lapotre_model_2019} to help stabilize channel banks, many of Earth’s Pre-Cambrian rivers may have been braided rivers. These ancient rivers formed massive fluvial conglomerates \citep{church_fluvial_1980} and flowed across the surface of the Earth for the majority of its history. Unfortunately, deposits from this period are rare $-$ Titan’s rivers may therefore offer an active analogy of these now largely lost rivers. Observations by \textit{Dragonfly} could also easily confirm whether alluvial channels exist on Titan and what their channel form is, especially if provided with observations of the grain size of material comprising the channel bed and banks. Documenting the gravel-sand transition \citep{lamb_grain_2016,dingle_gravel-sand_2021} by \textit{Dragonfly} would also be invaluable, providing a desperately needed new data point to resolve why such a stark transition exists in Earth’s alluvial rivers \citep{dingle_gravel-sand_2021}.  

The long-duration but decadal storms \citep{faulk_regional_2017,battalio_global_2021,battalio_interaction_2022} that may source Titan’s flood-stage flow discharges also require careful re-examination as to the concept of the bankfull discharge on Titan \citep{leopold_hydraulic_1953,wolman_frequency_1960}. On Earth this long-standing assumption that a 1$-$2 year flood discharge sets a channel's morphology is beginning to be questioned \citep{naito_can_2019}; Titan should be included in such discussions given that it may represent an extreme end-member of terrestrial rivers. Finer resolution observations of Titan's rivers by an orbiter, along with opportunistic in situ observations of active flow in a Titan river by \textit{Dragonfly}, would allow us to quantitatively assess the recurrence of flows in Titan's rivers, providing important new insight into Titan's climate. Further, despite being largely arid \citep{mitchell_climate_2016}, Titan's rivers may behave differently than arid rives on Earth \citep{pfeiffer_century_2021}, as the long-duration storms may fill up subsurface aquifers and sustain flows long after the storms cease. 

Though it is still difficult to discern any given river's precise geometry with the current \textit{Cassini} data \citep{miller_fluvial_2021,birch_geometry_2022}, additional theoretical studies in anticipation for \textit{Dragonfly} would still be important. The rates at which potentially icy sediment is transported off the Selk Crater rim would aid in \textit{Dragonfly}'s mission planning so as to find and sample icy, potential impact melt pond-sourced, material. Predictions of how a river's geometry may differ at the lander-scale would also aid in interpretations of any future landing site image data. Theoretical studies of the many dynamics that take place within a river would also be of value to any future Titan orbiter. Predictions of how Titan's materials and environments affect a river's observable geometry, what features may be present/absent on Titan, and how their geometry and scale may vary would all be of value as such mission concepts begin to be developed. Simply, application of the many tools and models used to understand river morphodynamics on Earth and Mars, though largely lacking data at the present, would still be of great value as we continue to explore Titan.  
\newline

\textit{6. \underline{Do Titan's ternary fluids produce new dynamics not encountered on Earth?}}\newline
Changes in the salinity and temperature of water drive circulation within Earth's oceans, and affect dynamics at freshwater and saltwater interfaces. Whereas saltwater and freshwater differ by $\lesssim$10\% in density, the potential temperature and pressure sensitivity of Titan's methane-ethane-nitrogen fluids can yield changes up to $\sim$20\% \citep{steckloff_stratification_2020}. Accordingly, there may be additional dynamic processes not usually encountered on Earth.   

Such dynamics may be especially prominent when a river meets a coastline. The composition of Titan's rivers may match that of the rain, dominantly methane-nitrogen \citep{graves_rain_2008}, while the seas will be comparably ethane-enriched (Figure \ref{Titan_bathy}). Depending on the temperature of the river, a river plume could either plunge if the river is colder than the sea, or be buoyant if the river were warmer. Bubbling due to nitrogen exsolution at the interface of the fluids \citep{farnsworth_nitrogen_2019,malaska_laboratory_2017} may further complicate river plume behaviors. Such dynamics also have significant implications for how sediment will be deposited along shorelines, if at all. Numerical investigations that consider the temperature evolution of Titan fluids as they flow across the landscape and within Titan's rivers are therefore needed. Laboratory investigations should also test the theoretical predictions of \citet{steckloff_stratification_2020} so as to constrain the magnitude of these density effects given what is known about Titan's current atmosphere \citep{mitchell_climate_2016,jennings_titan_2019}. Analog laboratory experiments that investigate how river plumes behave and how they deposit their sediment depending on density and viscosity contrasts would also aid in our understanding of such dynamics.   

Titan's ternary fluids would also affect how fluids flow to the seas, how they circulate within the seas, and the magnitude to which sea circulation feeds back on Titan’s atmosphere. Wind-driven currents have been shown to be capable of moving fluid from sea to sea, and can be as high as 5 cm/s \citep{tokano_wind-driven_2015,tokano_sun-stirred_2016}. Predicted tidal amplitudes, corrected for expected crustal deformation, range from $\sim$1 centimeter for Ontario Lacus to $\sim$1 meter for Kraken Mare \citep{tokano_simulation_2010}, and are capable of moving fluids and sediment \citep{schneck_coasts_2022}. Finally, the temperature-dependent viscosity and density variations of Titan's fluids \citep{steckloff_stratification_2020} could lead to density-driven flows. Does denser methane supplied from the surrounding landscapes, along with evaporation from more southern portions of the sea generate currents of equal, or even greater magnitude than those from winds or tides? The relative feedbacks between these various processes therefore offer a rich environment to apply the quantitative techniques of oceanography to Titan, with implications for understanding its present-day climate. 

Understanding how fluids circulate within Titan's seas will also be an important step toward understanding one of the most striking revelations from \textit{Cassini}’s altimetry measurements: the composition of the lakes and seas lack significant volumes of ethane or higher-order hydrocarbons (Figure \ref{Titan_bathy}). Ethane was predicted to be present in bulk within Titan's seas from photochemical models \citep{lunine_does_1993,krasnopolsky_photochemical_2010} and early VIMS detections at Ontario Lacus \citep{brown_identification_2008}, and so where and how the ethane is stored on Titan therefore remains one of the biggest mysteries in Titan science \citep{nixon_titans_2018}. 

Most often, methane and ethane clathrates are discussed as an efficient means to both resupply Titan’s atmospheric methane \citep{tobie_episodic_2006,choukroun_stability_2010}, and sequester ethane \citep{mousis_equilibrium_2014,vu_rapid_2020}. Clathrate storage is a promising mechanism given that methane favorably and instantaneously swaps with ethane in the subsurface \citep{mousis_equilibrium_2014}. Yet surface ethane will both circulate within the seas and must diffuse to increasingly deep clathrate layers within Titan's crust, the timescales and feasibility of which have not been quantified. Further, while clathrates are speculated to be present on Titan given they are thermodynamically favored \citep{carnahan_dynamics_2022}, we await measurements from \textit{Dragonfly} as to whether they do indeed exist. 

Titan's fluids are therefore unique from water on Earth, with density and viscosity variations potentially far more extreme than what we are familiar with. Though Titan's landscapes at \textit{Cassini}'s resolution look broadly like those on Earth, the potential added wrinkle of Titan's ternary fluids require new models and continued laboratory investigations. Image and topographic data from a potential orbiter could reveal the fine-scale nature of Titan's coastlines, the study of which would reveal the degree to which various currents are capable of modifying Titan's coasts \citep{schneck_coasts_2022,palermo_coasts_2022}. Long-baseline observations by an orbiter may also even be capable of detecting changes, including the transport of sediment along coastlines, river plumes, and active currents that roughen the sea surface. A more complete understanding of Titan's surface fluid dynamics is then the first step needed to understanding Titan's hydrologic and climate cycles more broadly, with the implications touching on many of the broader questions listed here.   
\newline

\textit{7. \underline{What can Titan's coastlines reveal about climate variations and long-term methane inventories?}}\newline
Earth's coastlines serve as the interface between the land surface and oceans. Coastlines are home to river deltas and other sedimentary landforms that together retain some of our planet’s best-preserved archives of both current and past hydrologic cycles \citep{posamentier_eustatic_1988}. Deltas are locations where numerous physical and chemical processes intersect, as liquids of different compositions meet and material eroded from the landscape (both solid and dissolved) is concentrated in a small area. Accordingly, how coastlines respond to climate variations, how deltas form and evolve, and what information they retain are areas of active research. 

Titan's coastlines should be important for many of the same reasons that they are on Earth, yet they remain rather understudied. Principally, Titan’s missing deltas represent a large knowledge gap in how fluid and sediment are interacting when a river meets a coastline. As such, the fate of the mass eroded from Titan’s landscapes remains unclear. Sediment transport calculations suggest that Titan's rivers should convey sediment to the coasts if provided a supply from upstream \citep{birch_geometry_2022}, while \textit{Cassini} should have no problem identifying deltas or other coastal landscapes \citep{birch_detection_2022}. Possible solutions to the question of Titan’s missing deltas then fall into two main categories: (1) deltas like those we see on Earth do not form (or rarely form) on Titan because of differences in materials, dynamics, and coastal conditions between the two worlds, or (2) there are many deltas on Titan, but the characteristics of the deltas make them difficult to identify – Titan’s deltas may simply “look” different. Distinguishing between these two scenarios, and decoding Titan’s record of tectonics, erosion and climate, requires new investigations of the processes that occur when rivers enter Titan’s seas.

The ubiquitous pits around Titan’s large lakes and seas (Figure \ref{Titan_seas}) may be akin to sinkholes. On Earth, sinkholes can form in similar places near coastlines and can even be found underwater (i.e., blue holes). As these features require drastic changes in sea level and a dissolvable substrate, studies on the formation of such features on Titan could reveal new information on the rates of sea level change and the composition of materials coating the seafloors. Titan’s coastlines, whether they are rocky or composed of sediment, are also likely affected by wind-driven waves \citep{palermo_coasts_2022}. To what extent, however, and whether erosion and/or transport of material along Titan’s coasts can outpace sea level variations is not entirely known. Given that Titan’s fluid density and viscosity may be highly sensitive to temperature \citep{steckloff_stratification_2020}, and that multiple sedimentary materials with different densities may be present, the diversity of sedimentary processes on Titan’s coasts may also be dizzying. One can imagine beaches of different density sediments (e.g., ice clathrate, benzene, co-crystals, etc.) clustered along even just the one shoreline of Ontario Lacus (Figure \ref{Titan_seas}), an exciting speculation confirmable only with finer resolution image data.

Just as on Earth, Titan is thought to also experience climate-driven sea level variations. Where waves deposit their energy and are capable of eroding bedrock and/or transporting sediment \citep{palermo_coasts_2022,schneck_coasts_2022}, and where deltas form both depend on the shoreline position. Sea level variations will therefore have a significant impact on Titan's shoreline morphology, the magnitude of which has yet to be fully quantified. Accordingly, the opportunity exists to combine observations of Titan's present-day shoreline morphology with numerical models of shoreline evolution to understand rates of sea level changes and, in turn, constrain Titan's climate \citep{black_estimating_2012,palermo_coasts_2022,schneck_coasts_2022}.  
\newline

\textit{8. \underline{How do Titan’s sharp-edged depressions form?}}\newline
Finally, Titan's SEDs appear to be a landform that is unique to Titan, as they seem to lack any known terrestrial (or planetary) analog \citep{hayes_topographic_2017}. A consistent understanding of how Titan’s SEDs form also remains elusive. The lack of inflow or outflow networks into these lakes imply that formation does not involve significant fluvial processes. Fluvial processes would also be unable to readily remove material from within the basins to create their observed flat floors \citep{hayes_topographic_2017}. The topographic correlation between nearby filled SEDs and empty SED floor elevations also supports subsurface communication and fluid flow \citep{hayes_topographic_2017}. The possibility for subsurface flow between Titan’s lakes was first discussed by \citet{hayes_hydrocarbon_2008} and has inspired some authors to suggest karstic dissolution as the preferred method of SED formation (e.g., \citet{cornet_dissolution_2015}). This mechanism is consistent with uniform scarp retreat and can explain the topographically-closed nature of the SEDs. Sublimation of a volatile substrate that underlies a more resistant surface layer can also work, although a lack of significant temperature gradients \citep{jennings_surface_2016,jennings_titan_2019} makes sublimation-based erosion ineffective in the current climate \citep{cottini_spatial_2012}. However, neither dissolution nor sublimation processes can explain the raised rims that bound many of Titan's SEDs \citep{birch_raised_2019}. 

Both dissolution and sublimation-based mechanisms also require a large deposit of volatile material to be concentrated at the poles, as water-ice is both insoluble in liquid methane/ethane and thermodynamically stable in Titan’s atmosphere \citep{lorenz_erosion_1996,perron_valley_2006}. This volatile material would have had to be delivered to the poles by processes other than uniform deposition of photolysis products, which are only predicted to generate $\sim$10 meters of solid material over the age of the solar system \citep{krasnopolsky_photochemical_2010}. Volatile caps could have been produced at Titan’s poles if acetylene and other volatile organics have been preferentially transported poleward over geologic time \citep{birch_geomorphologic_2017}, the exact thickness and formation timescale of which remain unconstrained. 

Alternative hypotheses proposed for SED formation include crater lakes \citep{lorenz_crater_1994}, cryovolcanism \citep{wood_morphologic_2020}, and explosive maar-like mechanisms \citep{mitri_possible_2019}. The planform morphology and dense clustering of SEDs argues against an impact-based hypothesis while explosive volcanism and/or caldera collapse does not explain the raised rims. With that said, cryovolcanic processes and/or other intrusive doming processes may act as initial seeding mechanisms for SED formation, with subsequent growth determined by other mechanisms. While periglacial processes can produce terrestrial features that are morphologically similar to Titan’s SEDs (e.g., pingos), there is no available material that can provide the required freeze-thaw cycles and wetting by hydrocarbon liquid does not appreciably change the volume of water ice clathrate \citep{choukroun_stability_2010}. 

Distinguishing between these formation models and/or developing new formation models for the SEDs therefore promises to reveal new information about Titan's materials and long-term climate evolution. Numerical formation models that can explain the many contrasting observations \citep{hayes_topographic_2017} would be of obvious value, which also require support from laboratory investigations into the properties of potential polar materials.  
\newline

\section{Summary}
Titan's coupled climate-landscape system is complex and varied, being both similar to and distinct from Earth's own system. While Titan's chemistry is different (methane vs. water, ice and organics vs. silicates), its landscapes and climate nevertheless present a useful and unique analogue to those studying the evolution of the terrestrial planets, especially Earth. At the kilometer-scale, \textit{Cassini} revealed fluvial and lacustrine landforms with a similarly rich diversity and strikingly similar forms to their terrestrial cousins. Despite our data limitations, these landscapes tell the story of a complex yet inherently familiar set of processes that act over range of timescales. Disentangling these will take decades of research. 

While Earth will always remain the primary locale for studies of landscape-climate feedbacks, study of Titan's fluvial and lacustrine landscapes can help probe the underlying physical principles and processes that drive climate evolution of Titan, Earth, and hydrologic worlds within and outside our solar system more generally. When future exploration by \textit{Dragonfly} and orbiter(s) provide regional-scale interrogation of fluvial landforms at the meter scale we will, without a doubt, see a whole new world. Continued exploration \citep{mackenzie_titan_2021} and interrogation of Titan should therefore revolutionize our understanding of how rivers and lakes evolve and interact with Titan's climate system. In doing so, we will be able to address far-reaching questions that touch on how planetary climate and hydrologic cycles are maintained through time, and how planetary hydrologic cycles are impacted by non-terrestrial materials. When ultimately provided with new data, and with the many tools constructed in the interim, Titan therefore promises to open wide the doors to the outer solar system for the many terrestrial and martian geologists, oceanographers, geomorphologists, and climate scientists. 

\section*{Acknowledgements}
The chapter was improved due to thoughtful reviews from Juan Lora and Jason Barnes. This work was supported by the Heising-Simons Foundation (51 Pegasi b Fellowship to S.P.D.B).

\bibliography{Titan_Hydrology_Book_Chapter}{}
\bibliographystyle{aasjournal}

\end{document}